\documentclass[12pt,preprint]{aastex}

\usepackage{natbib}
\slugcomment{ }

\shorttitle{Comparison between AR 12192 and AR 10486}
\shortauthors{Jain et al.}

\begin{document}


\title{Probing Subsurface Flows in Active Region NOAA 12192 - Comparison with NOAA 10486   }


\author{Kiran Jain, S.C. Tripathy and F. Hill}
\affil{National Solar Observatory, 3665 Discovery Drive, Boulder, CO 80303, USA}
\email{kjain@nso.edu, stripathy@nso.edu, fhill@nso.edu}


\begin{abstract}

Active Region (AR)  12192 is the biggest AR observed in solar cycle 24 so far. 
This  was a long-lived AR which survived  for
four Carrington rotations (CR) and exhibited several unusual phenomena. We measure the
 horizontal subsurface flows in this active region in multiple rotations
 using the ring-diagram technique of local helioseismology and the Global Oscillation Network Group
(GONG+) Dopplergrams, and investigate how different was the plasma flow 
in AR 12192 from that in AR 10486.  Both regions
produced several high M- and X-class flares but had different CME productivity. 
Our analysis suggests that these ARs had unusually large horizontal
flow amplitude with distinctly different directions. While 
 meridional flow  in AR 12192 was poleward  that supports
the flux transport to poles,  
it was equatorward in AR 10486.   Furthermore, there was a sudden increase
 in the magnitude of estimated zonal flow in shallow layers in AR 12192 during the X3.1 flare, however, 
it reversed direction in AR 10486 with 
X17.2 flare. These flow patterns produced strong twists in 
horizontal velocity with depth  in AR 10486   that persisted throughout the disk passage 
as opposed to AR 12192, which produced a twist only after the eruption of the X3.1 flare that
disappeared soon after. Our study indicates that the sunspot rotation combined with the re-organization
of magnetic field in AR 10486 was not sufficient to decrease the flow energy even after several 
large flares that might have triggered CMEs. Furthermore, in the
absence of sunspot rotation in AR 12192, this re-organization of magnetic field  
contributed significantly to the substantial release of flow energy  after the  X3.1 flare.
    
\end{abstract}



\section{Introduction}

During the course of each solar activity cycle, hundreds of ARs 
form and decay with time. These regions are responsible for 
producing energetic events, e.g., flares and coronal mass ejections (CMEs), 
thus, known as the major drivers of space weather. Observations show 
that a significant number of CMEs originate from flares \citep{Yashiro06}.
There are further evidences that big ARs are generally formed during the declining
phase of the cycle and produce more flares and CMEs than in ascending and maximum activity periods. 
Many authors have investigated the 
possible relationships between flare properties and the CMEs physical
parameters \citep[][and references therein]{Compagnino17}, and also tried to determine the
flare characteristics that may lead to the CME production \citep[][and references therein]{Liu16}.
In general, big ARs do produce both flares and CMEs.
However, there are few examples where large regions did produce flares but no CMEs.

\begin{figure}   
   \centerline{
\includegraphics[scale=0.9]{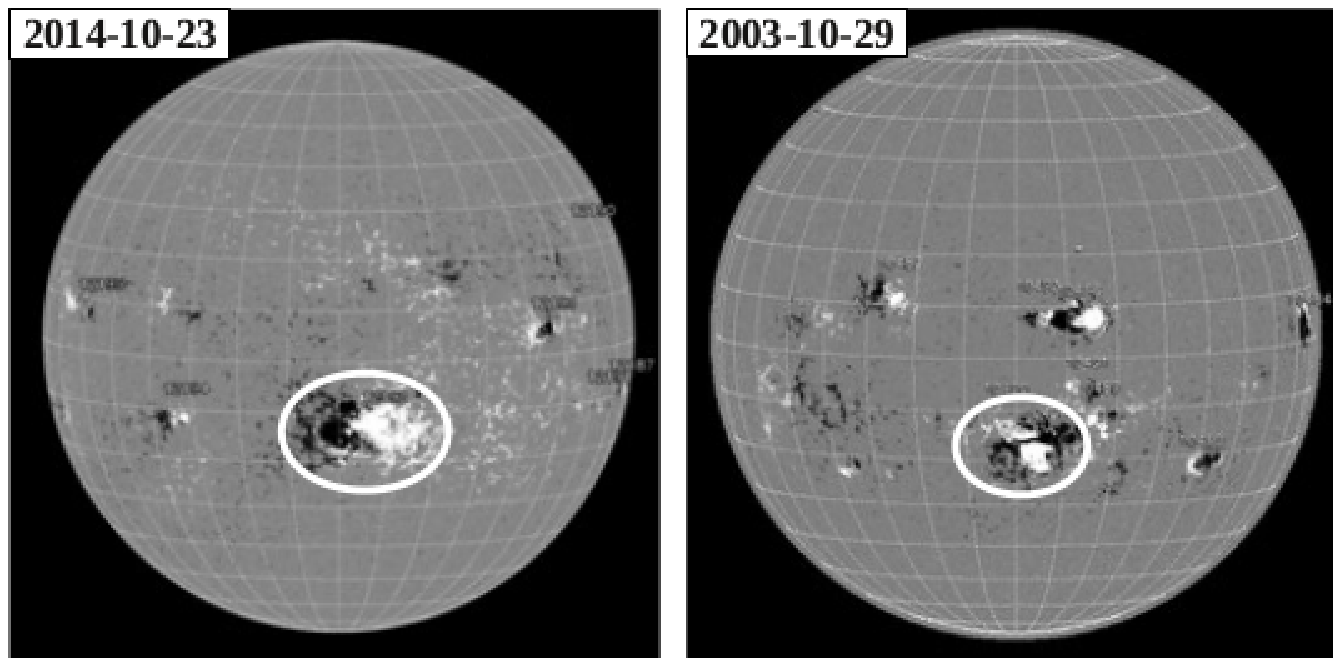}
}
   \centerline{
\includegraphics[scale=0.75]{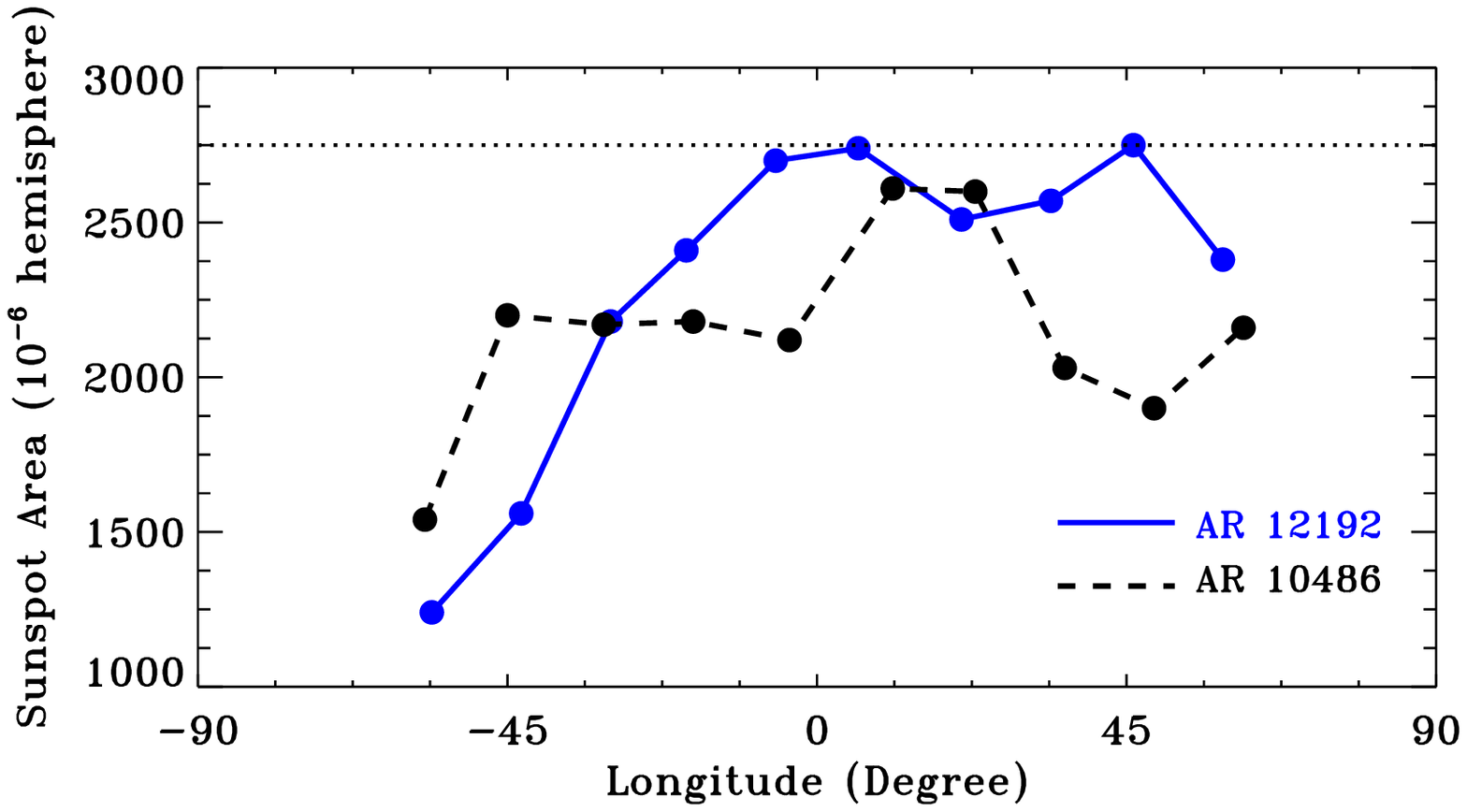}
}
            \caption{(Top): Magnetograms showing AR 12192 (left: from SDO/HMI) and AR 10486 (right: from SOHO/MDI) near the central meridian in Solar Cycles 24 and 23, respectively. These regions are marked by white circles. (Bottom):  Temporal evolution of sunspot area in both ARs during their disk passage  (Source: NOAA/SWPC). Dotted line indicates the maximum area acquired by AR 12192. 
}
   \label{ar_mag}
   \end{figure}

The most recent example is the NOAA 12192,  the biggest AR in last 25 years which
produced significant number of flares during its disk passage. However, only one small CME was 
reported to originate from this region. In total, there were total
73 C-class, 35 M-class and 6 X-class flares during 14 days of the disk passage
(ftp://ftp.swpc.noaa.gov/pub/warehouse/).
Among them, the most intense flare produced was X3.1
on 2014 October 24. In contrast to AR 12192, the biggest AR
of Solar Cycle 23, AR 10486,  produced 16 C-class, 16 M-class, 7 X-class flares
(including X17.2 flare on 2003 October 28) and several associated halo
CMEs (https://cdaw.gsfc.nasa.gov/CME\_list/).  In Figure~\ref{ar_mag}, 
we show sample magnetograms of  12192 (top left) and AR 10486 (top right)
around the same location on the solar disk but almost 11 years apart. 
The lower panel illustrates the variation of sunspot area in AR 10486 and 12192  as 
these pass through the Earth-facing side of the Sun. 
It can be seen that AR 12192 was much bigger in size than AR 10486.

Since the appearance of AR 12192, several studies  have been carried out to explore its properties
\citep{Chen15,Sun15,Jiang16, Liu16,Inoue16}. However,
these were confined to studying the magnetic conditions during the X3.1 flare with a focus
on understanding its CME-poor nature. \citet{Sun15} found a weaker 
non-potential core region in AR 12192 with stronger overlying fields, and smaller 
flare-related field changes when  compared with other major flare-CME-productive ARs 
11158 and 11429 of Cycle 24. By studying various magnetic indices, they further 
added that the large amount of magnetic free energy in AR 12192 did not translate 
into the CME productivity. The 
strong confinement from the overlying magnetic field responsible for the poor 
CME production was also  confirmed by  \citet{Chen15}.
 In a different study, \citet{Jing15} also compared the characteristics of the X3.1 flare of 
AR 12192 with the X2.2 flare of CME-productive AR 11158 in terms of magnetic 
free energy, relative magnetic helicity and decay index of these ARs. The major 
difference between them was found to lie in the time-dependent change of magnetic 
helicity of the ARs. AR 11158 showed a significant decrease in magnetic 
helicity starting 4 hours prior to the flare, but no apparent decrease in helicity was 
observed in AR 12192. By studying decay index, these authors also confirmed the earlier findings of 
\citet{Sun15} and  \citet{Chen15} that the strong overlying magnetic field was
mainly responsible for low-CME productivity in AR 12192. 
In order to provide more insight on this, 
\citet{Liu16} suggested that the large ARs may have enough current and 
free energy to power flares, but whether or not a flare is accompanied by a CME is 
related to the presence of a mature sheared or twisted core field serving as the seed of 
the CME, or a weak enough constraint of the overlying arcades. In a recent study, by
analyzing SDO/HMI, SDO/AIA and Hinode/SOT observations, \citet{Bamba17}
studied the X1.0 flare that occurred in the same region on 2014 October 25, which also
did not trigger any CME. They found that reversed magnetic shear at the 
flaring location was responsible for this 3-ribbon flare, and  the closed
magnetic field overlying a flux rope did not reach the critical decay index that prohibited the CME eruption.

Despite many studies describing the properties of AR 12192 above the solar surface, none has been published so far 
on the subsurface characteristics. There are no well-defined parameters 
from  helioseismic studies that favor the CME-poor or CME-rich nature of
any AR. However, in order to explore the subsurface signatures of eruptive events, 
\citet{Reinard2010} studied the temporal changes in subsurface helicity 
preceding AR flaring.  \citet{tripathy2008_CME}
also investigated the changes in acoustic mode parameters before, during, and after the onset of 
several halo CMEs. They found that the CMEs associated with a low value of magnetic flux regions
have line widths smaller than the quiet regions, implying the longer life-time for oscillation
modes. However, an opposite trend was found in the regions of high-magnetic field which
is a common characteristic of the ARs  \citep{Jain08}. 
Both these studies were based on  limited samples of ARs, hence a general
conclusion was not achieved.

In this paper, we study subsurface flows associated with AR 12192 and compare
with those in AR 10486. We  also investigate the differences between them in order
to understand the CME-poor nature of AR 12192. 
The paper is organized as follows; the observations of
AR 12192 in multiple CRs are described in Section~2. We briefly describe the
ring-diagram technique  in Section~3, and subsurface flows in AR 12192 and their 
comparison with AR 10486 in Section~4.  We discuss possible scenario supporting
different characteristics of both ARs in 
Section~5, and finally our results are summarized in Section~6. 

\section{Observations}

\subsection{Early observations of AR 12192 in EUV}
Two ARs, NOAA 12172 and 12173, appeared near the east limb in CR 2155. These regions
were formed on the backside of the Sun and were closely located.
 Continuous observations show that both regions survived till they reached the west limb, however, AR 12173
decayed within a few days after crossing the limb while AR 12172 went through rapid evolution.
The AR 12172 was later labeled as AR 12192 after its appearance on the east limb 
in CR 2156. To track the growth
and decay of AR 12172 and AR 12173, respectively, in particular on the invisible surface the the Sun,
we use ``all Sun" EUV Carrington maps  as described in \citet{Liewer2014}. 
These maps are produced by combining EUV frontside images from Atmospheric Imaging
Assembly (AIA) on board {\it Solar Dynamics Observatory (SDO)} with the backside 
images from  {\it Solar Terrestrial Relations Observatory (STEREO)}.  

\begin{figure}   
   \centerline{
\includegraphics[scale=0.68]{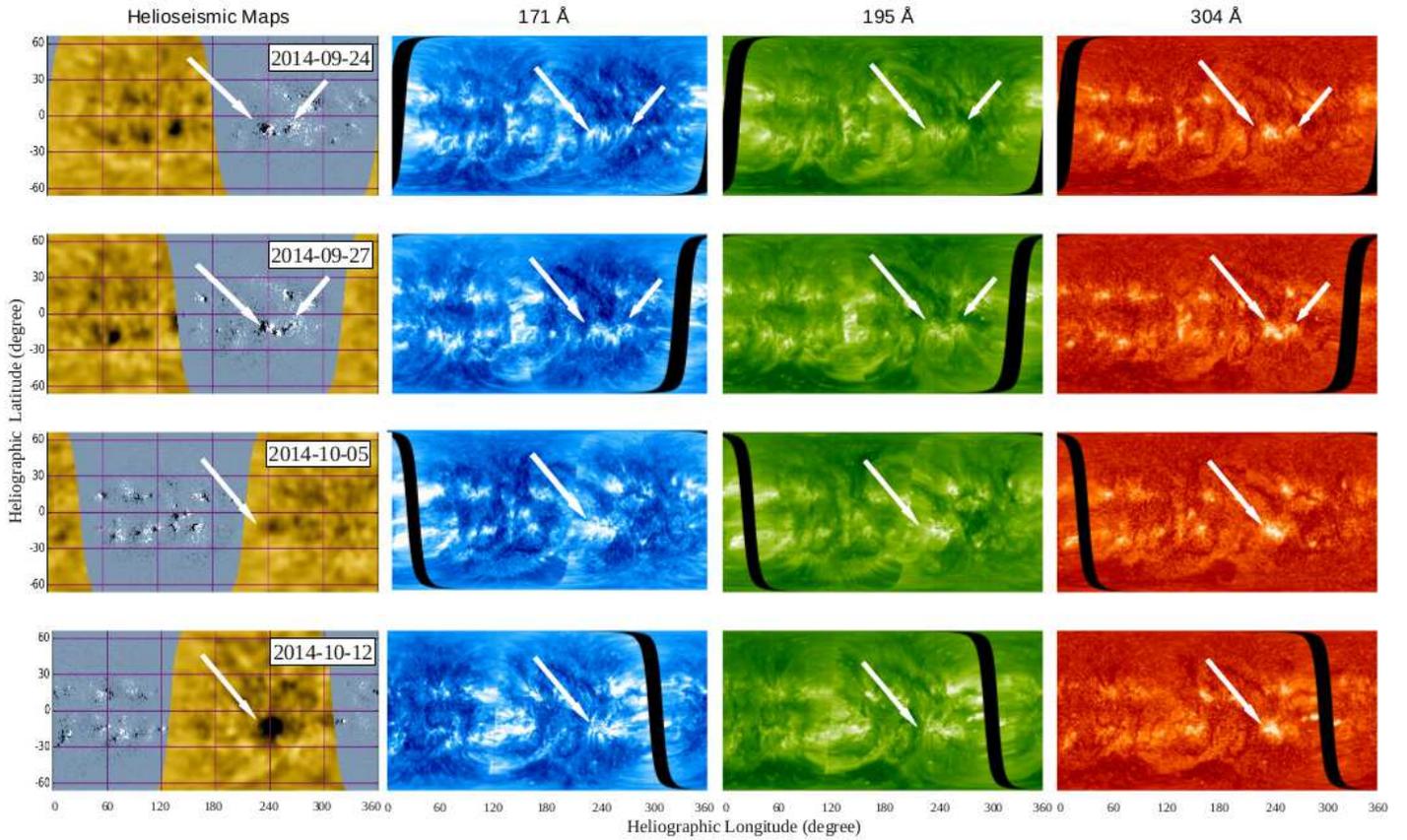}
}
            \caption{ (left) Composite "All Sun" Carrington maps of farside helioseismic images (yellow)
and frontside line-of-sight magnetograms (grey), and  corresponding maps in three EUV lines:  (2nd left column) 171\AA, (3rd left column) 195\AA~ and (right) 304\AA. The location of AR 12192 is marked by the large arrow in each
panel. Panels in first two rows have small arrows showing the location of AR 12173 which decayed after crossing the western limb. Following the standard practice adopted by the NOAA, 
AR 12192 was identified as AR 12172 in CR 2155. 
}
   \label{EUV}
   \end{figure}

 In Figure~\ref{EUV}, we show sample helioseismic and corresponding EUV maps in 195\AA, 171\AA ~and 304\AA, ~and for four days. The helioseismic maps are the composite images of the far- and near-side solar 
hemispheres, where the near hemisphere is represented by the line-of-sight magnetic field and the
farside by an image showing sound wave travel time variations.
 The locations of shorter travel times are darker regions indicating the  locations of high-magnetic field 
concentration on the farside. The top two rows provide maps when AR 12172 and 12173  were
present on the front side, and bottom two rows for those days when they crossed the west limb. Similar to
travel time maps,  the high magnetic field regions in EUV images can easily be determined  
in the form of brightening in these images. 
It should be noted that communications with the STEREO Behind spacecraft were interrupted on 2014 Oct 1
that hampered the required telemetry of the data. Nevertheless, a few images from STEREO do exist
that indicate the decay of AR 12173 between Sept 27 and Oct 5. A closer inspection of the images 
on Sept 24 (Top row) clearly
shows two well-developed ARs in both photospheric magnetogram (last column) and
EUV observations while in 2nd row, showing observations for Sept 27, we notice dimming in the
brightness of one AR. This kind of reduction in EUV brightness is
generally referred to as the decaying phase of the AR or the depletion of magnetic
field strength. We further notice that only one
AR was present in EUV images on Oct 5 indicating that 
AR 12173 was completely decayed by this time. Although there is a gap in the STEREO observations between
Oct 5 and Oct 12, the helioseismic maps strongly support the rapid growth of a strong
magnetic field region on the farside which is discussed below. This region could not be further tracked 
in EUV due
to the interrupted communications with STEREO Behind spacecraft.

\subsection{AR 12192 in multiple Carrington rotations - helioseismic imaging}
Direct and continuous observations of the visible surface and the helioseismic far-side imaging
suggest that the AR 12192 was a long-lived active
region. A careful examination of these images clearly indicates that the AR survived 
at least for four CRs and went through various evolutionary phases. Although this is
the biggest region of the current solar cycle till date, it did not produce 
the largest flare of the cycle or any CME. 
Since NOAA assigns new numbers to all ARs whenever these appear on the east limb irrespective 
of their reappearance after completing a solar rotation, the far-side images are crucial to ascertain the 
life  of such ARs.  The reliability of far-side images of ARs was tested in a series of papers by
\citet{Liewer2012, Liewer2014} where  helioseismic farside maps were compared with backside images of the Sun
from NASA's STEREO mission.  They found that approximately 90\% of the helioseismic 
active-region predictions matched with the activity/brightness observed in EUV at 
the same locations.  It should be noted that despite providing reasonable information on the existence of
ARs in the non-earth facing side of Sun, the current method of helioeseimic imaging
has a limitation on sensing strong signal near the limb. However, the signal becomes stronger 
when the ARs move towards the far-side central meridian. 

\begin{figure}   
   \centerline{
\includegraphics[scale=0.56]{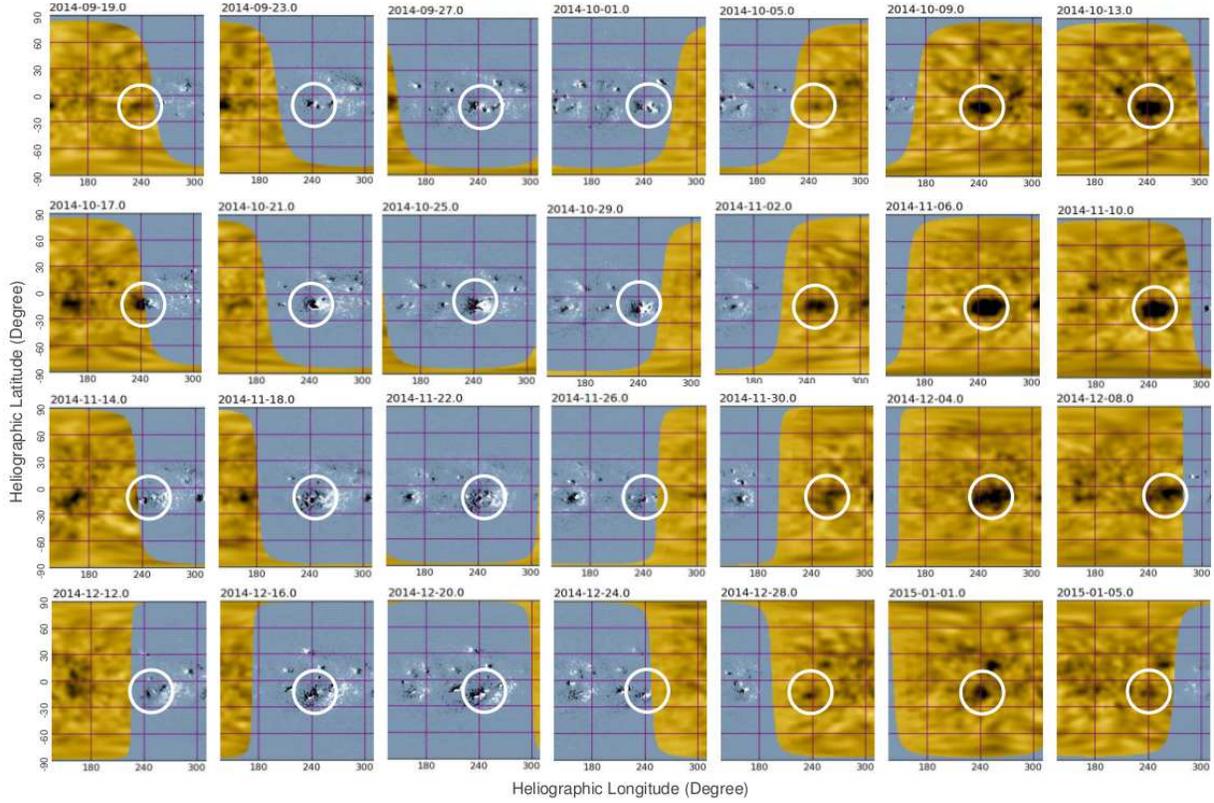}\\
}
            \caption{Far- and near-side images of  AR 12192 in four consecutive CRs: CR 2155 (Top row),
CR 2156 (2nd row), CR 2157 (3rd row) and CR 2158 (bottom row). Different
NOAA numbers were assigned to the AR in each CR.  These images
are produced by the farside pipeline of SDO/HMI and are available at http://jsoc.stanford.edu/data/farside/. 
   The existence of AR
12192 is marked by white circle in each image. Note that the AR disappeared after
crossing the western limb in CR2158. Gray portions in images display the front-side line-of-sight
magnetic field while yellow regions illustrates the far-side helioseismic maps. The farside maps produced 
by the NSO/NISP pipeline using GONG Dopplergrams (available at  http://gong2.nso.edu/products/)
also display similar results.  Dates corresponding to these maps are given at the top
of each panel. 
}
   \label{fs}
   \end{figure}

  In Figure~\ref{fs}, we show the images of AR 12192 in four consecutive CRs (CR 2155 - 2158)
by including images of both visible and invisible surfaces. Here we shown images for every fourth day with each 
row corresponding to AR's appearance in the next CR. The region of
interest is marked by the white circle. Daily helioseismic maps of far side
are available at http://jsoc.stanford.edu/data/farside/ using HMI and http://gong2.nso.edu/products/ 
using GONG observations. The evolution of AR 12192 in multiple CRs is 
summarized in Table~\ref{table1}.

\section{Data and Technique}

We utilize 1k $\times$ 1k continuous Doppler
images from GONG at a cadence of 60 s. Although better spatial resolution space observations
 from MDI on board SOHO during the 
Dynamics run from 1995 -- 2010 (1k $\times$ 1k at a cadence of 60 s) and 
from HMI on board SDO (4k $\times$ 4k at a cadence of 45 s)
are available  since 2010, the GONG 
observations provide a unique advantage in this work. 
 All these instruments use photospheric lines; the GONG and MDI  observations are in the \ion{Ni}{1} 6768\AA~line 
as opposed to HMI observations in \ion{Fe}{1} 6173\AA~line. Although the inference of subsurface
flow should not be influenced by the spectral line used in observations \citep{Jain06a},
 the different spatial resolution may introduce some bias on the results 
mainly due to the sensitivity to different mode sets for inversion \citep{Jain13}. In addition,
every instrument has some type of systematic effects, thus the use of consistent
data for both ARs as well as quiet regions will minimize instrument
related bias, if any. Furthermore, the GONG observations will allow us to use the same quiet region values
from the deep minimum phase between cycles 23 and 24 as reference.  It is worth mentioning
 that there have been changes in the GONG instrumentation over the period of this
analysis, however these were limited to adding new capabilities only, e.g., the modulator upgrade
in 2006 to obtain high-cadence magnetograms and adding H-$\alpha$ filter in 2010 at a 
20 s cadence for space weather studies, thus the
Doppler observations were not affected.  Therefore, the GONG network provides unique consistent data set of
continuous high-cadence and high-spatial resolution Dopplergrams for last one and half solar cycles
that can be used to study long-term solar variability.

We use the  technique of ring diagrams  \citep{Hill88} to study the subsurface plasma flow
in AR 12192 in multiple CRs.
 This method has been used in numerous studies to investigate the acoustic mode 
parameters beneath ARs \citep{Rajaguru01,Tripathy12, Rabello-Soares16} as well as long-term
and short-term variations in near-surface flows \citep{Komm15a,Greer15,Bogart15,Howe15, Tripathy15,Jain15}. 
It is widely understood that the strong concentration of magnetic field alters the
behavior of acoustic waves by absorbing their power. As a result, the inferences in
these regions often encounter with large uncertainties.
In a recent study,  the credibility of the ring-diagram method was tested by comparing 
near-surface flows with the surface flows in three ARs 
calculated from ring-diagram and the local correlation tracking methods, respectively   \citep{Jain16}.
The authors reported a good correlation between these two quantities. Studies further
show that the uncertainties in inferences tend to increase towards the limb  \citep{Jain13a}, hence
 our analysis is restricted to within $\pm$45${\degr}$ from the disk center.

\begin{figure}   
   \centerline{
\includegraphics[scale=0.9]{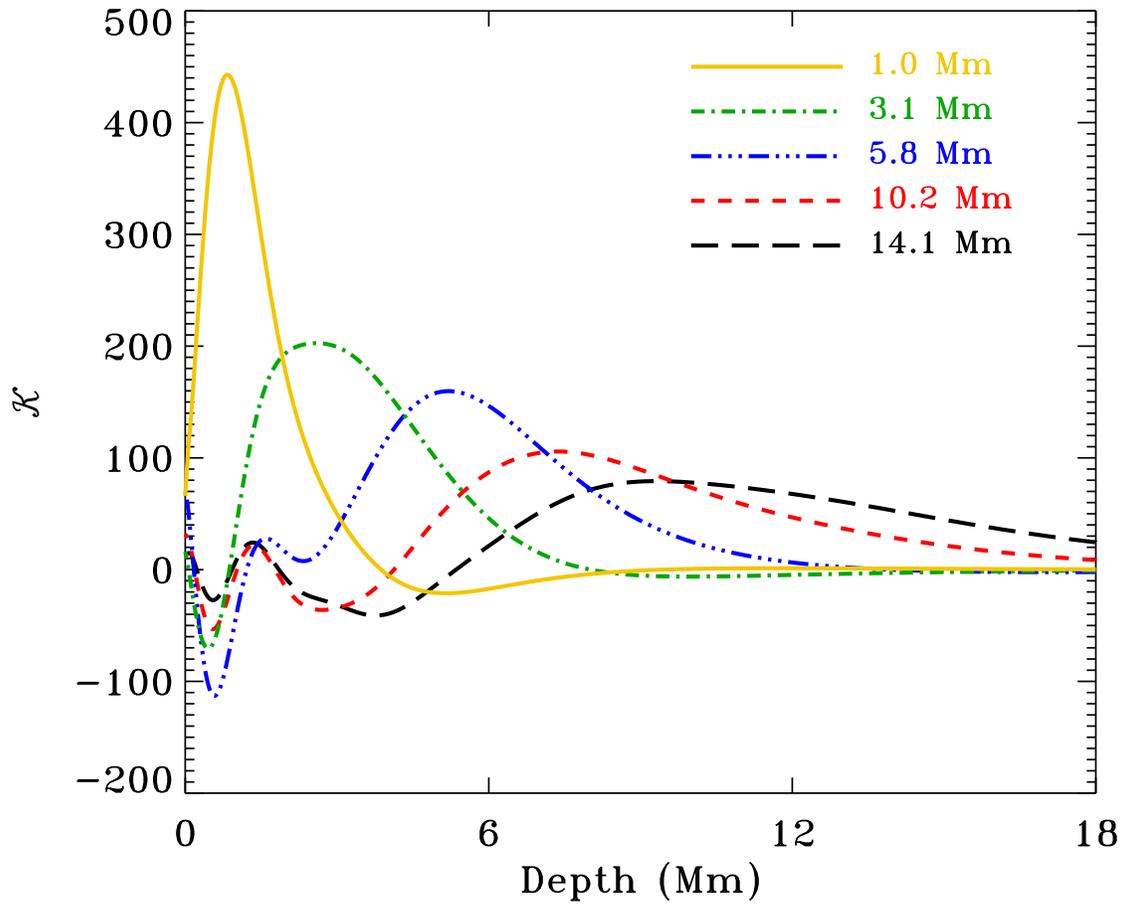}\\
}
            \caption{Representative averaging kernels used in the RLS inversion as a function of depth. The centroids
of these kernels are located at depths mentioned in the figure. The kernels become less localized with 
the increase in centroids' depths.
}
   \label{kernels}
   \end{figure}

 In the ring-diagram method, high-degree waves propagating in localized areas over the solar 
surface are used to obtain acoustic mode parameters in the region of interest. 
In this work, we track and remap the regions of $15^{\degr} \times 15^{\degr}$ (128 $\times$ 128 pixels
in the spatial directions)
for 1664 min (referred to as 1 ring day)  at the surface rotation rate \citep{snod84}.  
The tracking rate of each  central
latitude in these regions is of the form cos($\theta$)[$a_0 + a_2$sin$^2(\theta)  + a_4$ sin$^4(\theta)$], 
where $\theta$ is  the latitude and the coefficients 
$a_0$ = 451.43 nHz, $a_2$ = $ -$ 54.77 nHz, and $a_4$ = $ -$ 81.17 nHz. We
then apodize each 
tracked area with a circular function and a three-dimensional FFT is applied in 
both spatial and temporal directions to obtain a three-dimensional power spectrum that is fitted  
using a Lorentzian profile model \citep{Haber02},
\begin{eqnarray}
P (k_x, k_y, \omega) & = & {A \over (\omega - \omega_0 + k_xU_x+k_yU_y)^2+\Gamma^2}  + {b \over k^3} 
\end{eqnarray}
where $P$ is the oscillation power for a wave with a temporal frequency ($\omega$) and
the total wave number $k^2=k_x^2+k_y^2$. There are six parameters to be fitted:
two Doppler shifts ($k_xU_x$ and $k_yU_y$) for waves propagating in the orthogonal 
zonal and meridional directions, the background power ($b$), the mode
central frequency ($\omega_0$), the mode width ($\Gamma$), and the amplitude ($A$).
Finally, frequencies along with the fitted velocities ($U_x$ and $U_y$) and formal errors
are used as input to a regularized least square  (RLS) method to estimate depth dependence of 
various components of the horizontal velocity (zonal: $V_x$ and meridional: $V_y$).
While the zonal component provides an estimate of the flow in east-west
direction, the meridional component is for the north-south direction. 
 The depth dependences of mode kernels used in the inversion are shown in 
Figure~\ref{kernels}. It can be seen that the kernels sharply peaks near the surface
due to the availability of large modes for the kernel construction, however
as the centroids of the kernels increases in depth, the kernels become broader
and the inversion's depth resolution becomes poorer.                 

In order to investigate the subsurface variability with the evolution/decay of ARs,
we use quiet region values
as a reference and subtract them from the AR values in various time samples.
Our analysis is restricted to radial order $n$ = 0 -- 6 in 
the frequency range of 1700$\mu$Hz $\le \nu \le$ 5200$\mu$Hz. 
 It should be noted that we have used custom patches
processed through the latest version of the GONG ring-diagram pipeline (version 3.6.1).
Although both active regions studied in this paper were located 
close to the multiple of $7^{\degr}.5$
 in both directions (the grid spacing used in ring-diagram pipeline), the
 standard products provided different locations for
 AR 10486 on the solar disk from those for AR 12192. Hence, for the direct comparison between
 these ARs, we have customized the time series in AR 10486 which means the start 
and end times for each set are different from the standard products. 

\subsection{Quantitative estimates of magnetic activity}

 In our study, we also quantify the magnetic activity in various regions by calculating the  Magnetic 
Activity Index (MAI)  in each sample using the line-of-sight 
magnetic field from the magnetograms. Here we convert magnetogram data to absolute values, 
average over the length of a ring day and apodise them into circular areas to match the size 
of the Dopplergram patches used in the ring-diagram analysis. 
 We use both SOHO/MDI and GONG magnetograms to calculate  MAI.   Since the entire
 period under study is not covered by the MDI magnetograms alone, we use the 96 minute MDI 
magnetograms for 2003 and GONG magnetograms sampled every 32 minutes for 2008 and 2014.   
Lower threshold values of 50 and 8.8 G are used for MDI \citep{basu04} and GONG \citep{Tripathy15}, 
respectively, which are approximately 2.5 times the estimated noise in the individual 
magnetograms. In addition, we also correct the magnetic field data from space 
for cosmic-ray strikes. To generate a uniform set of MAIs, the MDI MAI values are scaled to GONG values 
using a conversion factor of 0.31, which was obtained by relating the line-of-sight 
magnetic field measurements between these two instruments using a 
histogram-equalizing technique \citep{Riley14}. 

\section{Results}

\subsection{Systematics and the selection of quiet regions}

In a study of the subsurface characteristics of any active
region in multiple time samples, the first step is to eliminate the influence of systematic 
effects, if any. The most common error comes from the foreshortening that may arise due to
the change in heliographic locations of the tracked regions during the disk passage.  To 
overcome this effect, we calculate reference values of quiet regions at the same
heliographic locations as the AR  and subtract them from the
AR values as described in \citet{Jain12,Jain15}. As a result, we obtain residuals 
that provide estimates of flows due to the change in magnetism. Another 
problem is the selection of quiet regions  which may also influence
the inferences. \citet{Tripathy09} have examined the role of the choice of quiet regions 
in helioseismic studies and demonstrated that the results derived for the pairs of 
quiet and ARs were biased by the selection of quiet regions. They
further suggested that an ensemble average of quiet regions should be used in such studies. 

In previous studies, we have chosen quiet regions from the neighboring Carrington
rotations \citep{Jain12,Jain15}. However, for an unbiased comparison between ARs 12192 
(CR 2155) and 10486 (CR 2009), we use same quiet regions from the deep
 minimum phase between cycles 23 and 24 for both ARs.  
Note that this was the period of extremely low solar activity, hence the values obtained 
with reference to this period can be termed as ``true'' AR values.
Although, both ARs appeared in two different solar cycles, this selection of
quiet regions is supported by the following observational facts; (i) both ARs 
were observed in same phase of the year when the values of the solar inclination
angle ($B_0$) were comparable,
and (ii) both ARs were located around the same 
latitude in the southern hemisphere (Figure~\ref{ar_mag}). Thus the projection effect in both
cases can be uniformly removed by using the same quiet-region values. The
period of 2008 September thru December is used to determine the quiet-region values.  

 We  compute error-weighted averages of the MAIs for 13 ring days (approximate time for any region to
cross the disk) at each heliographic location. Since we are interested in 
studying the temporal variation in the subsurface properties of 
AR 12192 in four CRs
and $B_0$ varies from +7${\degr}$.0 (September) to $-$1${\degr}$.5 (December) during the period
of analysis, the quiet regions are selected in such a way that the values of MAI do not significantly change
from region to region.   Average
MAIs for quiet regions at 15 disk positions along latitude 15${\degr}$  for four
different ranges of $B_0$ are summarized in Table~\ref{table2}. The  $B_0$ values
chosen here correspond to those for the active regions analyzed in this paper.
The regions along the latitude 15${\degr}$ are separated by 7${\degr}$.5 and 
cover the solar surface from $60{\degr}$E to $60{\degr}$W.
It can be seen that the MAIs vary from 0.3 to 0.6 G and the standard deviation 
is within 18\% with largest values near the limb 
decreasing to  about 8\% near central meridian. 

\begin{figure}   
   \centerline{
\includegraphics[scale=0.9]{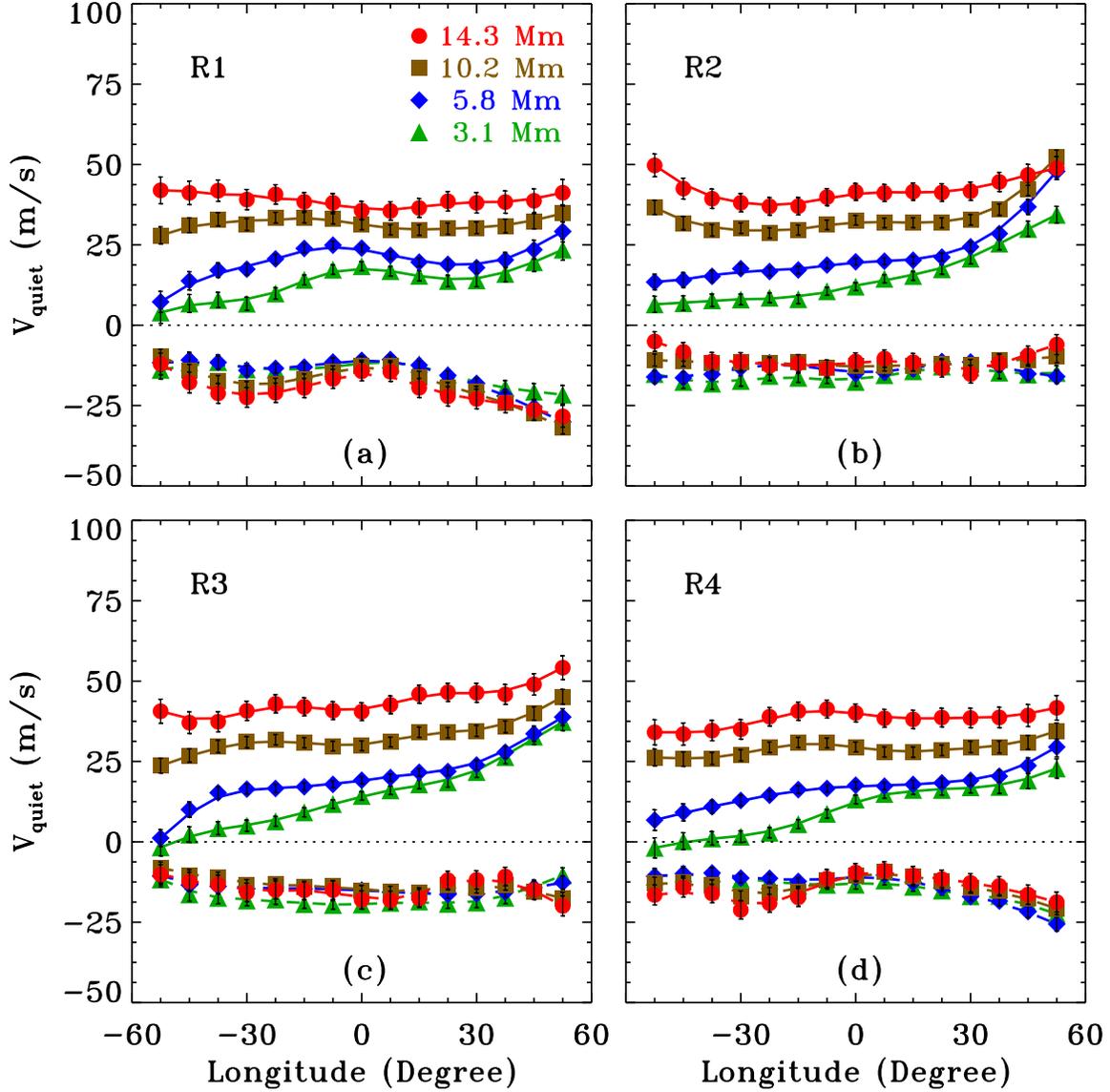}\\
}
            \caption{Variation of error-weighted averages of  zonal (solid) and meridional (dashed) 
components of the horizontal velocity as a function of disk location at 4 target depths during the minimum activity period in 2008. Symbols
represent the calculated values while lines represent for the 3-point running mean.
Each panel corresponds to a fixed range of $B_0$; 
(a) R1: 7$^{\degr}$.05 -- 6$^{\degr}$.51, (b) R2: 5$^{\degr}$.31 -- 3$^{\degr}$.99, (c) 
R3: 2$^{\degr}$.58 -- 0$^{\degr}$.91, and (d) R4: $-$0$^{\degr}$.69 -- $-$2$^{\degr}$.39.
These four ranges of  $B_0$ correspond to the data obtained from 2008 September to 2008 December.
}
   \label{flow_quiet}
   \end{figure}

\begin{figure}   
   \centerline{
\includegraphics[scale=0.85]{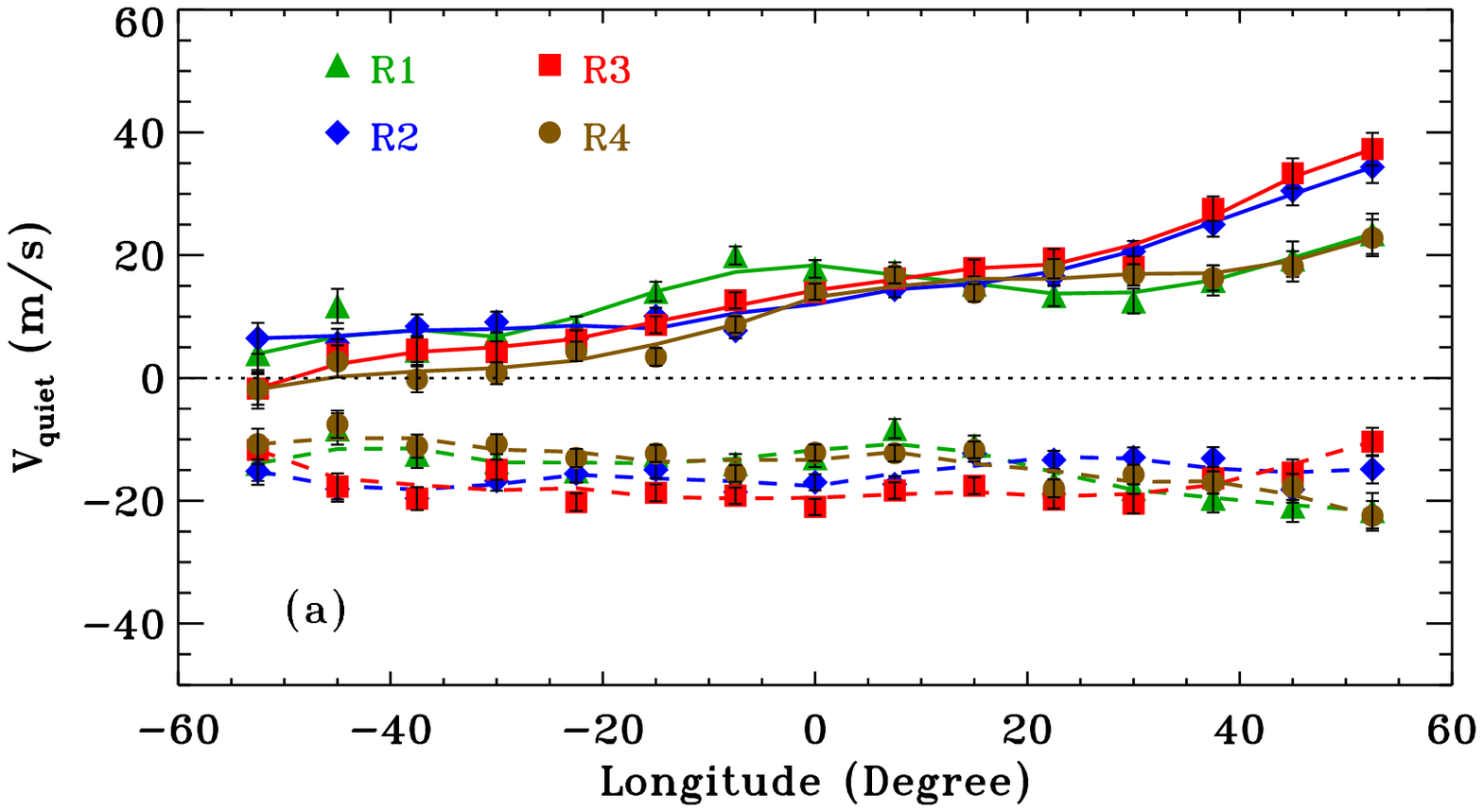}
}
 \centerline{
\includegraphics[scale=0.85]{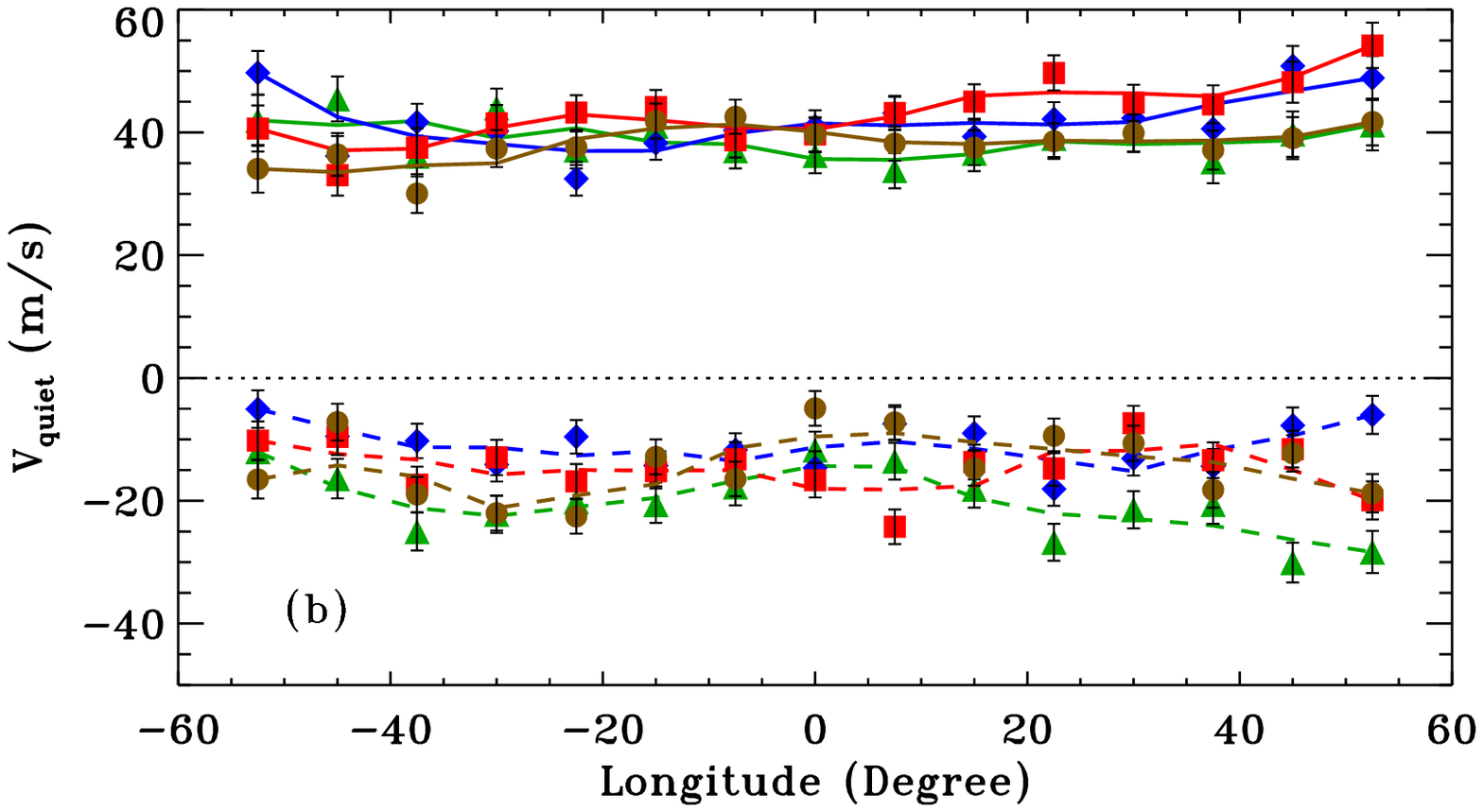}\\
}
            \caption{Variation of error-weighted averages of  zonal (solid) and meridional 
(dashed) components of the horizontal
velocity  as a function of disk location at fixed depths (a) 3.1 Mm and (b) 14.3 Mm. Symbols
represent the calculated values while lines are for the 3-point running mean of the calculated values
in different ranges of $B_0$; R1: 7$^{\degr}$.05 -- 6$^{\degr}$.51,
 R2: 5$^{\degr}$.31 -- 3$^{\degr}$.99, 
R3: 2$^{\degr}$.58 -- 0$^{\degr}$.91, and R4: $-$0$^{\degr}$.69 -- $-$2$^{\degr}$.39.
 These four ranges of  $B_0$ correspond to the data obtained from 2008 September to 2008 December.
}
   \label{flow_quiet1}
   \end{figure}

We further calculate the components of horizontal velocity corresponding to 
these quiet regions and compute error-weighted averages of 13 ring days at each location
 as for MAIs. As our selected regions have an overlap of 7${\degr}$.5, we smooth 
these components with 3-point running mean which are later used as reference in the active-region analysis. 
The longitudinal variation of these components across the disk is displayed  in Figure~\ref{flow_quiet}
where different panels represent the averages for different $B_0$  ranges at four target
depths below the surface, i.e. 3.1, 5.8, 10.2 and 14.2 Mm.  It is clearly seen that 
the magnitude of $V_x$ increases with depth which is in agreement with the observations 
where rotation rate is found to increase with depth in the outer 5\% of the Sun. 
There are some variations in the magnitude of $V_y$ as well, however these are within 1$\sigma$. Further, 
both zonal and meridional velocities show an east-west trend  but it is more prominent in the 
zonal component. Also, the shallower layers are more affected by this systematic, which decreases 
with depth. To highlight the  variation in east-west trend with time, we plot, in  Figure~\ref{flow_quiet1}, 
the variation of both components with disk location at depths 3.1 Mm and 14.2 Mm.
  We notice significant variations
 in these components with both time and depth. 
 These systematics  were also noticed by \citet{Zhao12} in the the time-distance
analysis where they found both travel-time magnitude and 
variation trends to be different for different observables. Though these authors did not
provide any physical basis for this effect, they suggested that these systematics must be
taken into account for the accurate determination of flows.  Later, \citet{Baldner12}  
explained this effect in terms of highly asymmetrical nature of the solar granulation
which results in  the net radial flow. This east-west trend was also studied by \citet{Komm15a} in flows from
the  ring-diagram analysis for both GONG and HMI observations,
where different data sets exhibited different systematics. They concluded that the trends 
could be removed adequately by selecting reference values from the same data source. Thus, 
in our study of the temporal variation of flows at different depths, we use reference values that
depend on both depth and the disk location of the active region.

\subsection{AR 12192 and subsurface horizontal flows}

\begin{figure}   
   \centerline{
\includegraphics[scale=0.8]{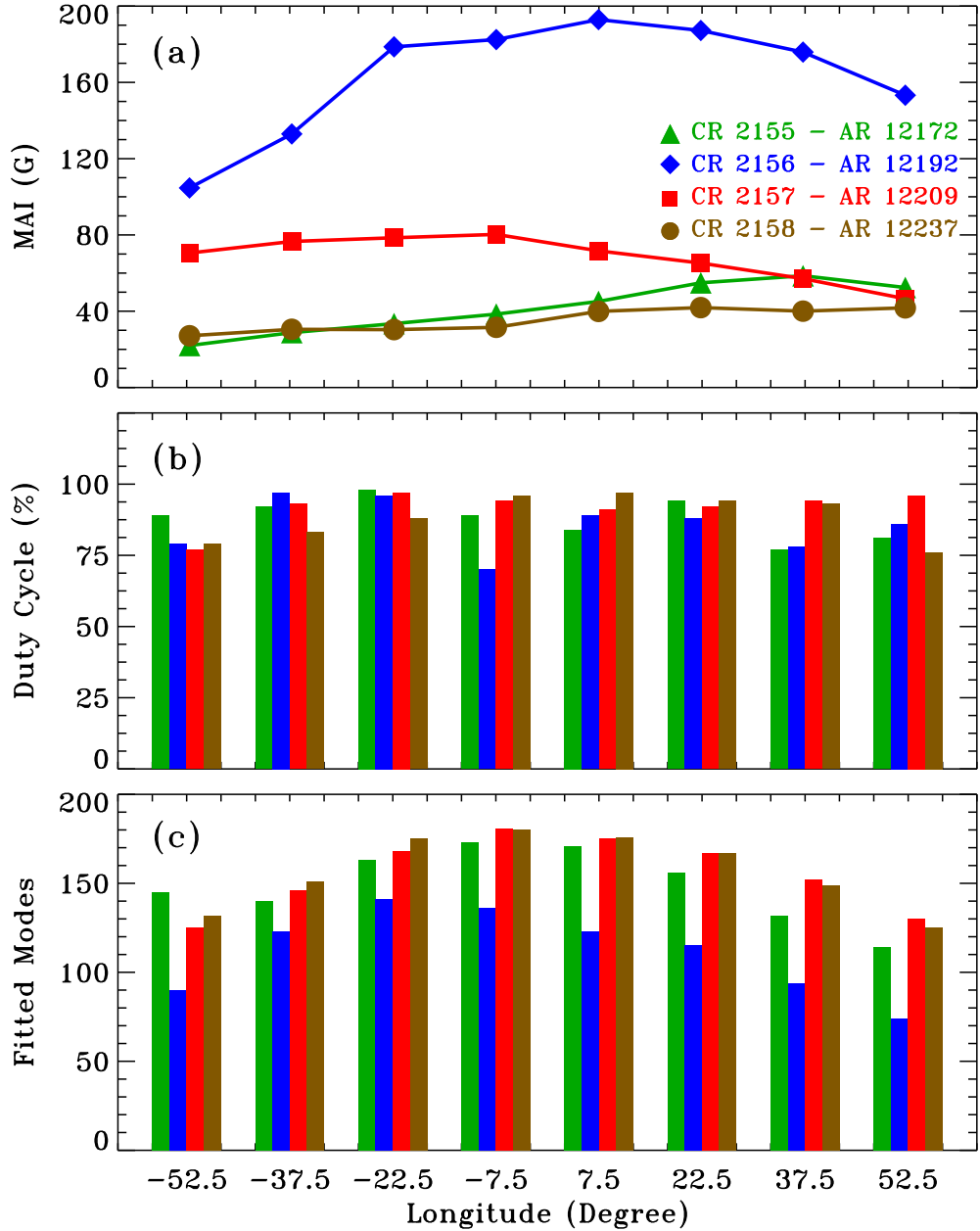}\\
}
            \caption{(a) Temporal variation of magnetic activity index (MAI) in AR 12192 in four 
consecutive CRs. Note that  different NOAA numbers are assigned to
the same region in different CR, (b) histogram showing the duty cycle corresponding to each region, and (c)the number
of fitted modes used in inversion to calculate the depth-dependence of flow fields. Note that the 
histograms are shown at fixed eight locations as described in text. 
}
   \label{mai2014}
   \end{figure}

Manifestation of strong magnetic fields is not only visible on the solar surface and beyond in the
 solar atmosphere, it also has significant influence on the subsurface properties. While most
acoustic mode parameters tend to change linearly with the increase or decrease in magnetic field 
strength \citep[][and references therein]{Tripathy15}, the velocity vectors have a
complex relationship. These are modified
in the presence of strong fields, however there are several other factors that may dominate amplitude
and the direction of velocity vectors.  In general, the topology of ARs plays an important role in
defining the velocity components. Also, moderately large number of modes
are needed in inversion for reliably calculating the depth dependence of flows.
The evolution or decay of the magnetic activity in AR 12192 in 
multiple CRs is illustrated in Figure~\ref{mai2014}.
Locations of the reference image (i.e. the image at the center of each timeseries)
 are given in Table~\ref{table3}. Since the Carrington longitude 
identifies the unique location of any region on the solar surface, we have made
sure that it does not change with time. The heliographic
location for this AR is 247${\degr}$.5 longitude and $-15^{\mathrm{o}}$ latitude. 
It can be seen that the magnetic field strength in this AR became stronger in 
the second rotation (CR 2156) which 
decayed to a moderate level in third rotation (CR 2157). It further decayed in the 
fourth rotation and the MAI values became comparable to those in the first rotation (CR 2155).
 
Among additional factors that may influence the inference of flow vectors,  the high
duty cycle is crucial in obtaining reliable flows. The large gaps in
observations or low duty cycle reduce the number of fitted modes that
have significant effect on the inverted flows and produce large errors.
Since the duty cycle in GONG observations varies from day-to-day, primarily due to the 
changing weather conditions at observing sites or to system maintenance/breakdown, the 
results may be  altered.  To authenticate our results, we also include 
duty cycles  in Table~\ref{table3} and Figure~\ref{mai2014} for each timeseries. 
It can be seen that these are  $\ge$ 70\% in all time series 
which is above the lower threshold set in earlier studies (R. Komm, private communication). 
The mean duty cycles for eight ring days in each CR are 88.0\%, 85.4\%, 91.8\% and 88.3\% with 
the variation between 1.2$\sigma$ to 2.3$\sigma$  within the CR.
Another factor is the heliographic location of the region  which
also influences the number of modes.
Thus, fitted modes are predominantly governed by all three factors; higher
magnetic field, low duty cycle and the larger distance from  disk center lead to a
lower number of fitted modes. It is evident from  the lower panel in Figure~\ref{mai2014} that 
the number of modes in CR 2156 at each location is relatively low though the duty cycles at 
some locations are
comparable to other CRs. This clearly demonstrates the effect of the significantly strong magnetic
field in CR 2156 on the acoustic modes, which reduces their amplitude and  as a result, produces large errors.
We estimate the  variation in number of 
modes is less than 2$\sigma$ in each CR
which is higher than the variation in duty cycle. 

\begin{figure}   
   \centerline{
\includegraphics[scale=0.9]{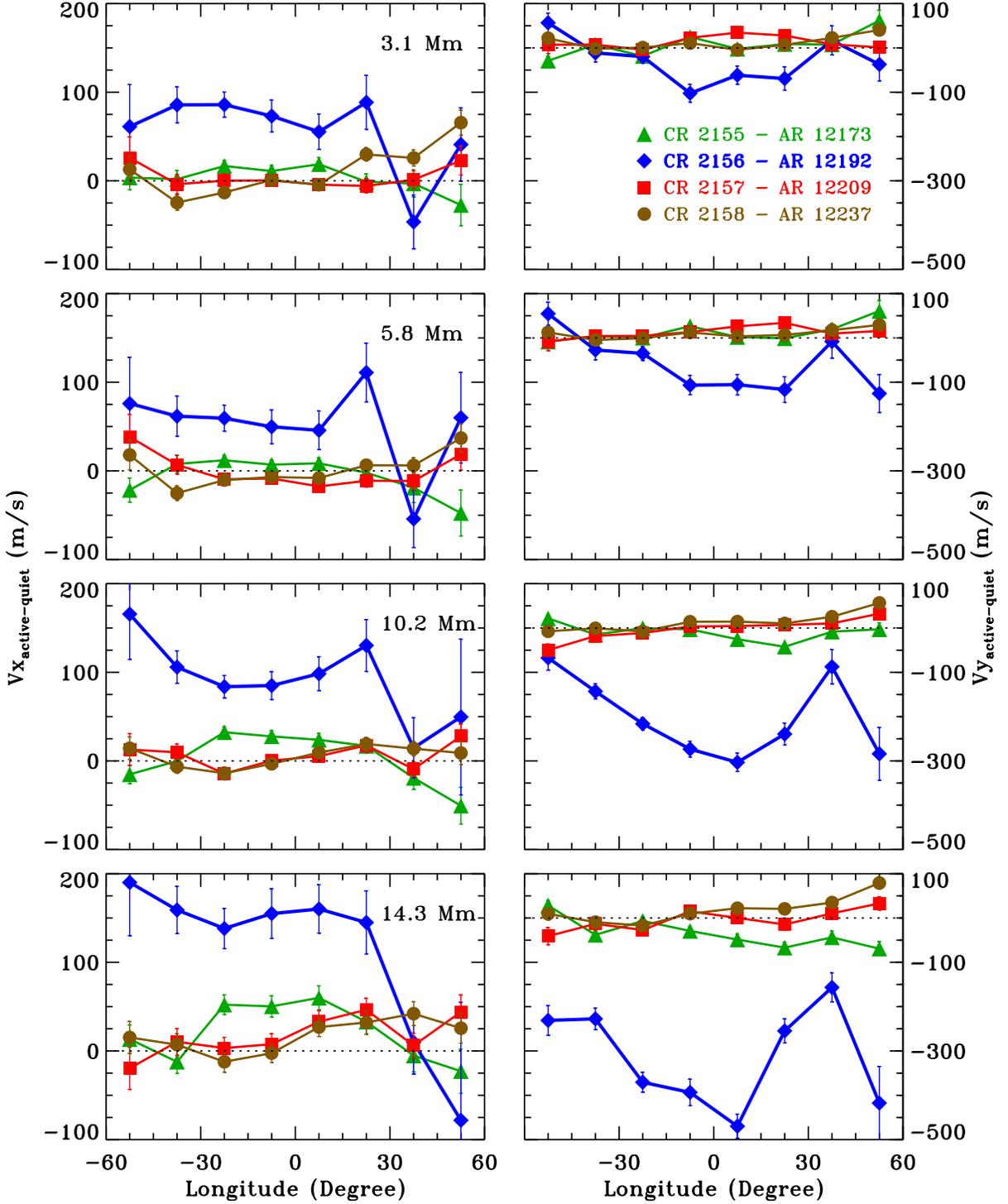}\\
}
            \caption{Temporal variation of zonal (left) and meridional (right) components of the horizontal velocity 
below  AR 12192 in four CRs at four depths below the surface. The uncertainties in 
velocity determination in all rotations except CR 2156 are smaller than the size of the symbols.
 Note that the velocity ranges are different for both zonal and meridional components.
}
   \label{flow_12192}
   \end{figure}

 Temporal variation of $V_x$ and $V_y$ for the AR  in each 
CR are displayed in Figure~\ref{flow_12192}. 
These values are obtained by 
subtracting the quiet Sun values at the same locations. In the absence of 
any high magnetic field region, the meridional component is typically poleward (negative values
in the southern hemisphere) that provide basis  for the magnetic flux transport from low 
latitudes to the poles, and the zonal component is in the direction of solar rotation. 
However, as mentioned above, their amplitudes as well as directions are modified  
due to the change in the characteristics of AR. 
We obtain maximum values of both velocity components in CR 2156 in comparison to other
three CRs.  It is further noted that the values of both  components are comparable (within 1$\sigma$)
 in shallow layers in all CRs except CR 2156, however these differ moderately in
deeper layers. We also find very large flows in deeper layers in CR 2156. Since the MAI values
are significantly large in CR 2156 (Figure~\ref{mai2014}), these could be responsible for the large flows and different behavior in this Carrington rotation.
We also notice exceptionally
large errors at $\pm$52${\degr}$.5 in CR 2156 that may arise due to lower number
of modes available for inversion as shown in  Figure~\ref{mai2014}. 
 
\subsection{Comparison with AR 10486}

\begin{figure}   
   \centerline{
\includegraphics[scale=0.6]{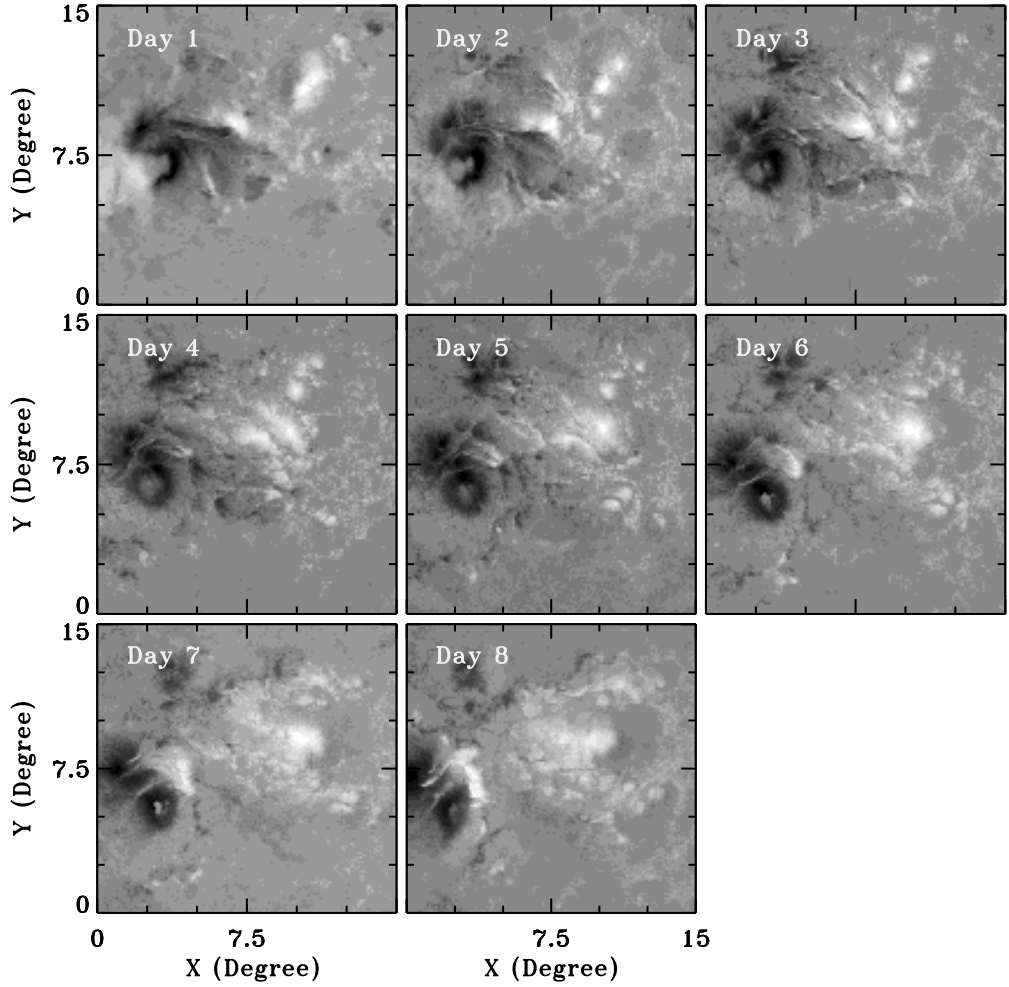}\\
}
            \caption{SDO/HMI magnetograms showing the evolution of AR 12192 in $15^{\degr} \times 15^{\degr}$ 
 tiles for eight consecutive ring days in CR 2156. The locations of these regions are described in  Table~\ref{table3}.
}
   \label{patches_2014}
   \end{figure}

\begin{figure}   
   \centerline{
\includegraphics[scale=0.6]{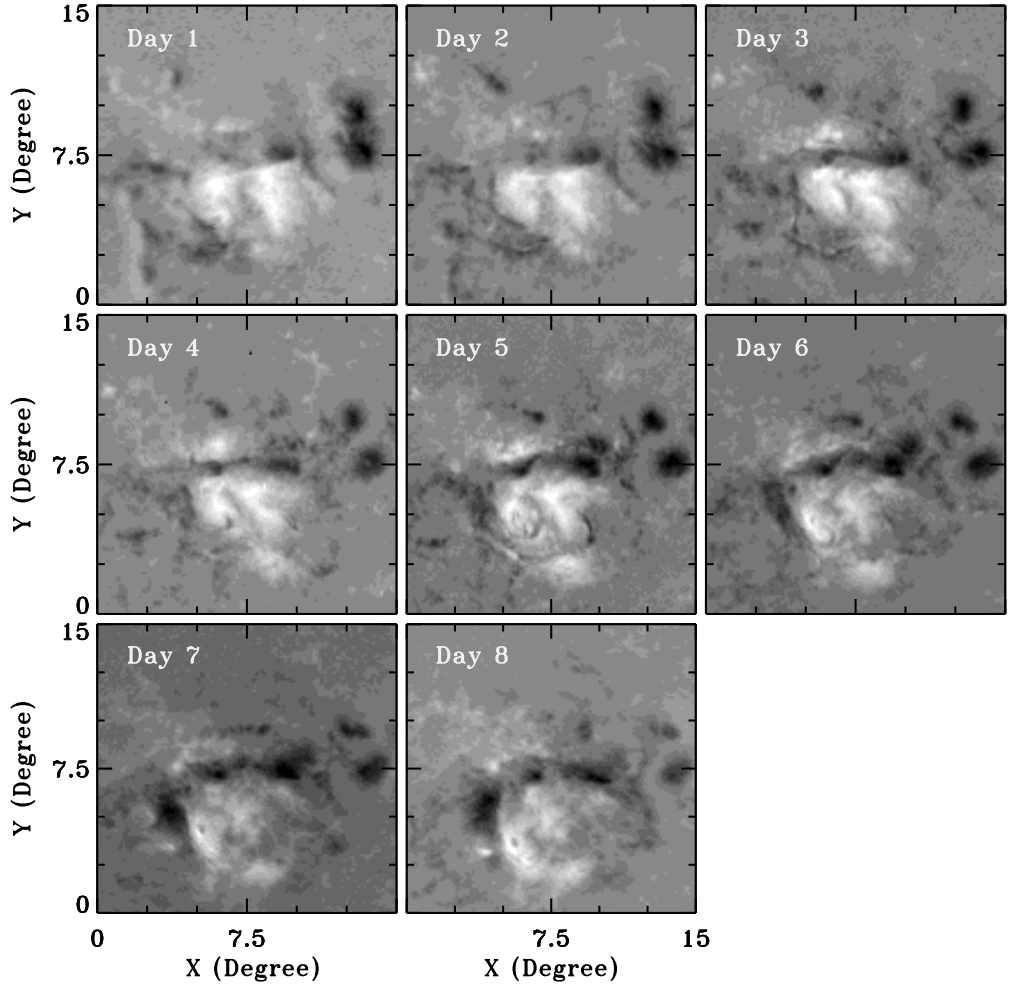}\\
} 
            \caption{SOHO/MDI magnetograms showing the evolution of AR 10486 in $15^{\degr} \times 15^{\degr}$ 
 tiles for eight consecutive ring days in CR 2009. The locations of these regions are described in  Table~\ref{table4}. 
}
   \label{patches_2003}
   \end{figure}

\begin{figure}   
   \centerline{
\includegraphics[scale=0.8]{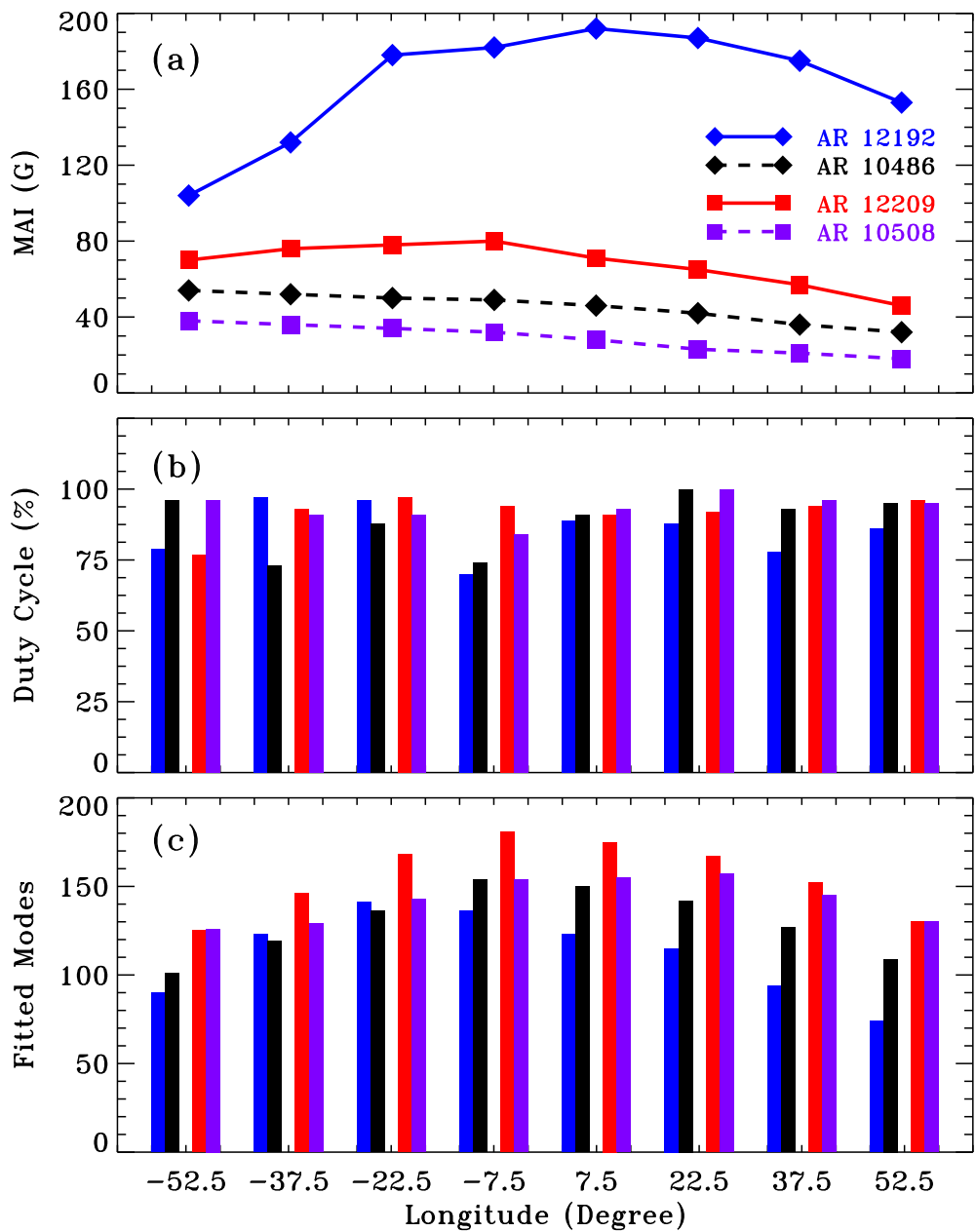}\\
}
            \caption{(a) Temporal variation of magnetic activity index (MAI) in AR 12192 (12209) 
and AR 10486 (10508)  for two 
consecutive CRs are shown by solid and dashed lines, respectively. Note that 
different NOAA numbers are assigned to these ARs  in next CRs and these
are in parenthesis, (b) histogram showing the duty cycle corresponding to each region, and (c) the number
of fitted modes used in inversion to calculate the depth-dependence of flow fields. Note that the 
histograms are shown at fixed eight locations as described in text, and AR 12192 (12209) and AR 10486 (10508)
are grouped together for comparison at each location. 
}
   \label{mai_2ar}
   \end{figure}

 The active region 10486 was a long-lived AR \citep{Irene07} that
produced several X- and M-class flares and associated CMEs in cycle 23.  The magnetic classification
of this region was similar to AR 12192,
 however the field strength was not as strong as it was in AR 12192. 
 There was a slow decrease in the magnetic field in AR 10486  in
the next CR while it decreased significantly in AR 12192. 
In order to compare the horizontal flows from both active regions, the position of
ARs in the tracked cubes is crucial. 
 If one region is off centered from the other,
the inferences of mode parameters and the flows will be biased by the area
covered. Further, quiet areas within the tile will modify the results
as calculated flows are averaged over the entire tile.
To reduce the biasing of these factors, as  mentioned in Section~3, both regions are 
tracked at the Snodgrass rotation rate which allows us to cover the same area in tracking
ensuring that  the ARs stay at the same location in the tracked cube. Sample images of the evolution 
of ARs 12192 and 10486 in the tracked cubes for
eight consecutive ring days are shown in  Figures~\ref{patches_2014} and \ref{patches_2003}, respectively.
These are the magnetogram regions near or at  the reference image which is the center of the each time series.
Since both ARs are large, it is clearly visible that these cover a
reasonably large area of the tracked regions with stable positions of ARs within the tiles.
The start and end of each time series  
representing   AR 12192 and AR 10486 at different disk locations are listed in Tables~\ref{table3} and
\ref{table4}, respectively. 
Temporal variations of MAI, duty cycle and the number of fitted modes in both regions
for  two consecutive CRs  are displayed in Figure~\ref{mai_2ar}.
As illustrated, the number of modes in AR 12192 and 10486 were comparable in the first two days and 
then differences started to arise which increased as ARs moved towards the west. 
These differences may introduce some bias in the flow estimates and the errors
are also large when different mode sets are available for inversion.
For convenience, in the following analysis, we will refer to 
regions in different CRs by their NOAA numbers. We, therefore,  compare the subsurface 
properties of AR 12192 with those of AR 10486, and AR 12209 with AR 10508 in the next rotation. 

\begin{figure}   
   \centerline{
\includegraphics[scale=0.9]{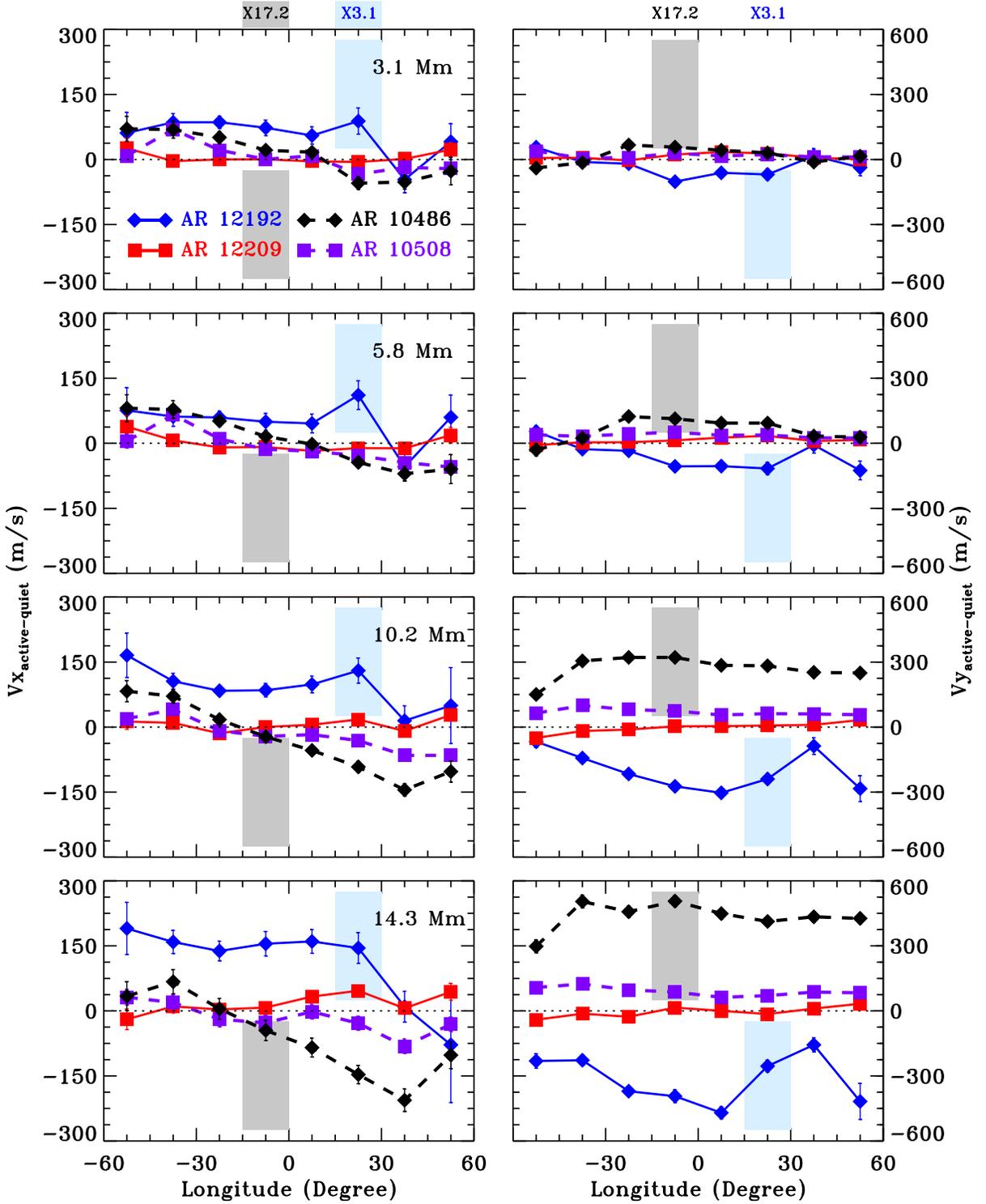}\\
}
            \caption{Temporal variation of zonal (left) and meridional (right) components 
of the horizontal velocity in  AR 12192 (solid) and AR 10486 (dashed) in two consecutive CRs at 
four depths below the surface.  The uncertainties in 
velocity determination are smaller than the size of the symbols.
 Note that the velocity ranges  are different for both zonal and meridional components. Colored boxes
show the locations of largest flares originated from these active regions.
}
   \label{flow_2003}
   \end{figure}

 Figure~\ref{flow_2003}  illustrates the daily variation of  zonal and  meridional velocities.
Our analysis reveals a number of similarities as well as differences between ARs 12192 and 10486. We
find that both ARs maintained significantly large flows during their disk passages but
 distinctly different directions.  Locations of X3.1 and X17.2 flares in ARs 12192 and 10486, respectively
are also marked in  Figure~\ref{flow_2003} and  also tabulated in Tables~\ref{table3} 
and \ref{table4}. The resultant 
amplitude of both components is found to increase with depth. While there is a small change in $V_x$ 
in AR 12192 several days before the X3.1 flare, the gradient is sharp in the case of AR 10486. In fact, 
direction was reversed in AR 10486 a day before the flare around 6 Mm producing a 
twist in the horizontal velocity, whose amplitude continued to increase with time. Interestingly, 
there was an increase in $V_x$ in AR 12192 at the time of  X3.1 flare which then decreased
 significantly with a small twist near the surface despite producing another X2.0 flare. 
Furthermore, the calculated meridional components showed a different trend; although there was a
small temporal variation in  AR 10486, the maximum values in AR 12192 in deeper layers
 were achieved a day before the  X3.1 flare with the decreasing trend after that. 

\begin{figure}
\figurenum{13a}   
   \centerline{
\includegraphics[scale=0.55]{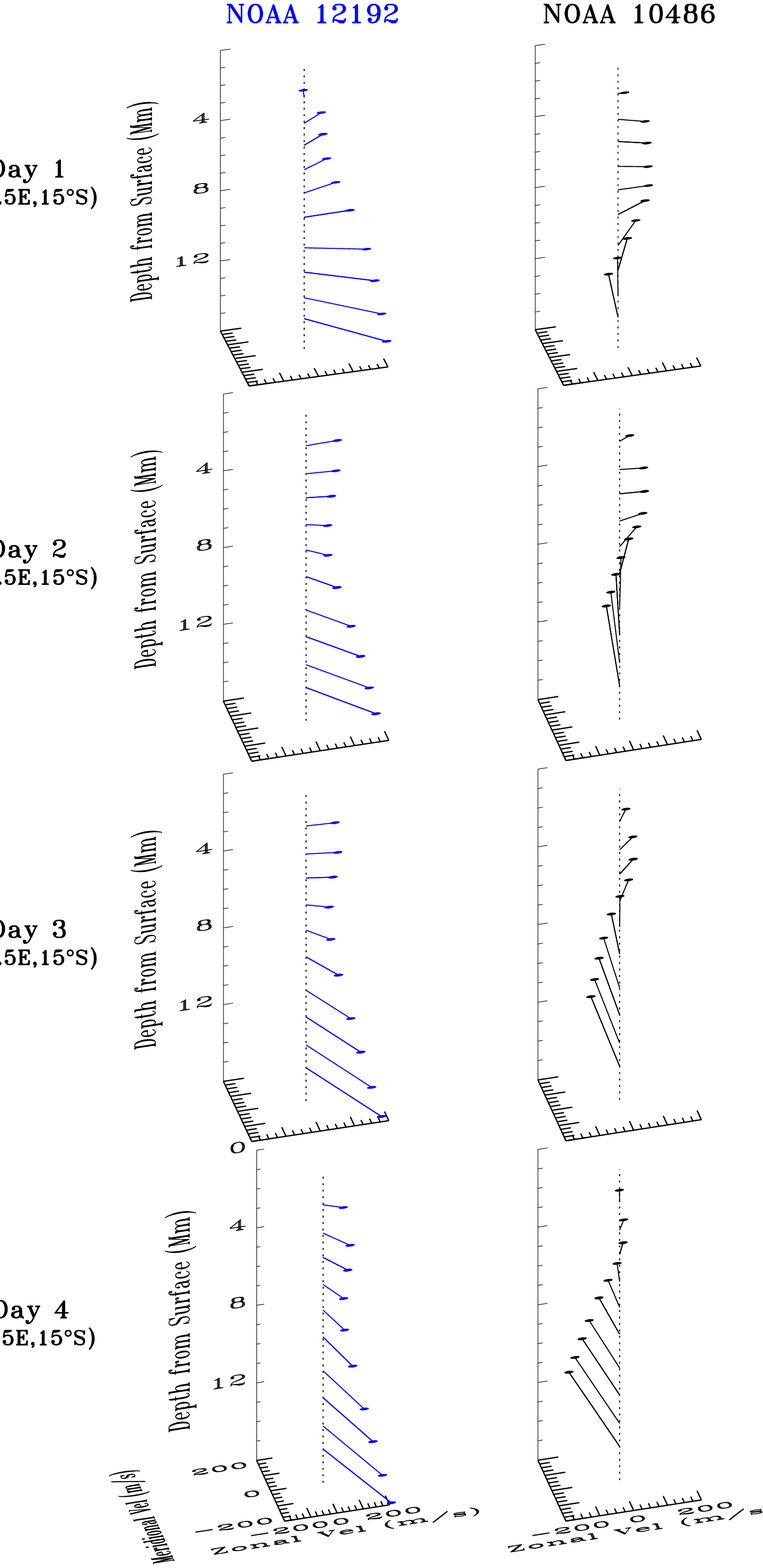}
}
           \caption{Depth variation of total horizontal velocity vectors in AR 12192 (left) and
AR 10486 (right) at  four disk locations in eastern hemisphere in CR 2156 and 2009, respectively.  Individual
zonal and meridional components are plotted in   Figure~\ref{flow_2003}. Note that the time
progresses downwards.
}
   \label{box_R2a}
   \end{figure}

\begin{figure}

\figurenum{13b}     
   \centerline{
\includegraphics[scale=0.55]{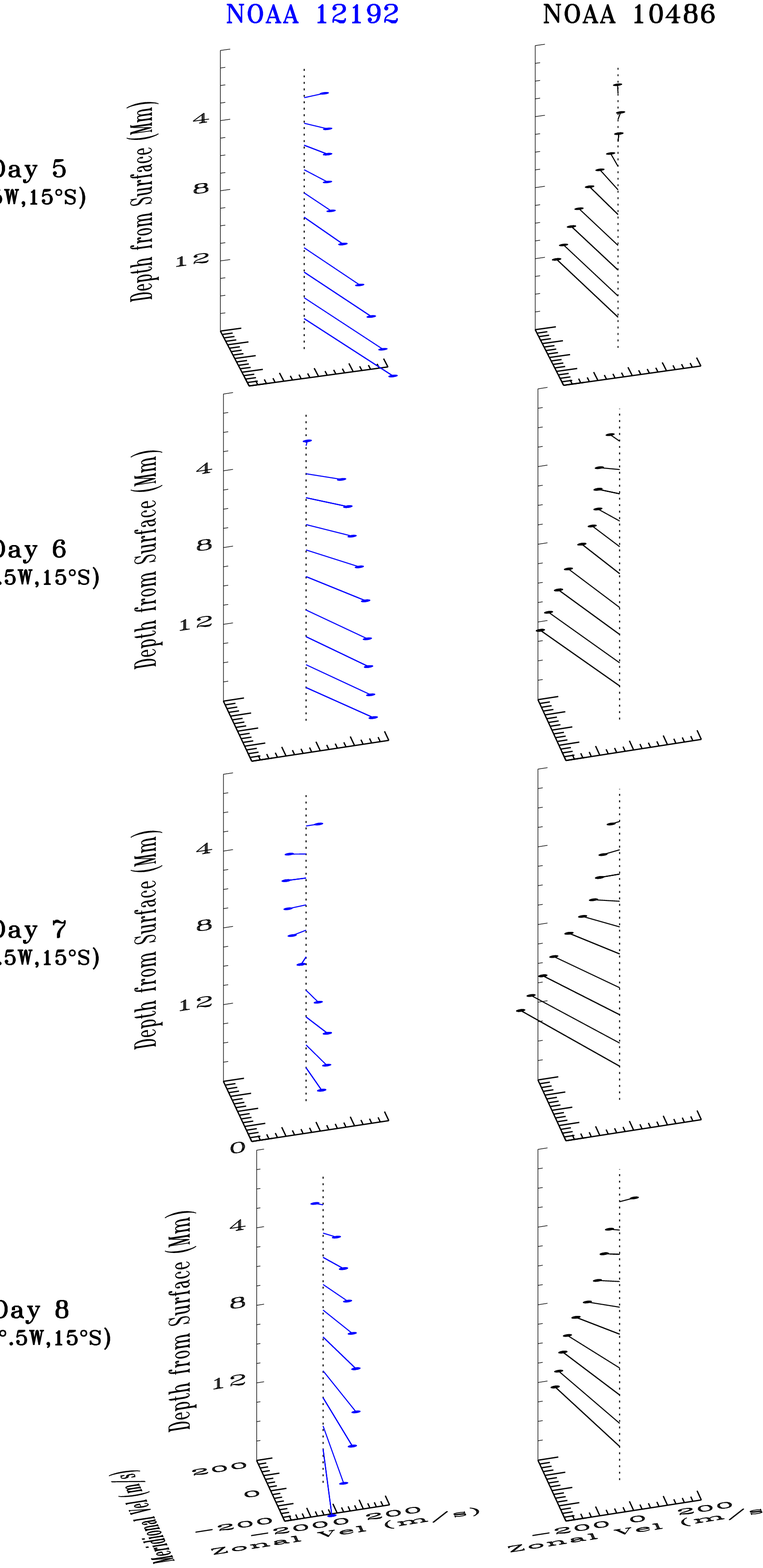}
}
           \caption{Depth variation of total horizontal velocity vectors in AR 12192 (left) and
AR 10486 (right) at four disk locations in western hemisphere in CR 2156 and 2009. Individual
zonal and meridional components are plotted in   Figure~\ref{flow_2003}. Note that the time
progresses downwards.
}
   \label{box_R2b}
   \end{figure}

 Figures~\ref{box_R2a} and~\ref{box_R2b}  illustrate the depth variation of estimated 
total horizontal velocity at eight disk locations where each panel represents 
an individual ring day. It is evident that there are strong twists in the flows for AR 10486
 and that the pattern persists for the first five days. Previous studies show that this 
kind of twist is generally associated with strong flares.  As indicated in Table~\ref{table4}, 
there were a series of X- and M-class flares from this AR before and after 
the major X17.2 flare on Day 4, which we find to be responsible for the observed twists for several days. 
Note that there was X8.3 flare on Day 8 that started near the end of time series
used in the analysis but continued beyond the end time. We find a slight twist
near the surface in this case too but a complete flow pattern associated with this
flare was not achieved.  Nevertheless, the twist appears in AR 12192 on Day 7 only 
after the X3.1 flare despite the fact that there were
other major flares before and after this event (see Table~\ref{table3}). 
Our analysis suggests that the twist disappeared thereafter when the region was 
centered at longitude 52$^{\degr}$5W. It is clear from  Figure~\ref{mai_2ar} that the number 
of fitted modes depends on the location of the region on the solar disk,
decreasing as the AR moves away from the disk center making the inferences noisier
with higher uncertainties. In addition, any magnetic field  within the region and 
the duty cycle influence this number. Further, the outer boundary of the reliable inferences 
depends on the spatial resolution of input Dopplergrams and the GONG resolution puts this limit at
 around 45$^{\degr}$ in either direction \citep{Jain13a}. However, we have plotted flows at 
52$^{\degr}$5W to understand their behavior after large flares, where we find the twist 
has disappeared and the values are expected to be biased by moderately large uncertainties. 
In next CRs, as evident from Figure~\ref{flow_2003}, 
 there was  a significant decrease in the flow amplitude in both ARs.
 
\section{The plausible scenario}

It is believed that the magnetic fields are generated in a thin layer at the
base of convection zone, known as the tachocline, hence the subsurface properties may
provide useful information on their evolution.  The techniques of local helioseismology
are widely used to obtain vital information from the convection zone beneath quiet as well active regions.
In last couple of decades, with the availability of high-resolution continuous
observations at high cadence, it has become possible to map plasma flow in the
convection zone and derive quantities that are directly linked with the 
measurements on the surface \citep{Komm15}.     
Further, \citet[][and references therein]{Komm12} carried out several studies on a large
sample of ARs to understand the overall behavior in their evolutionary and decaying
phases.  In addition, helioseismic studies describing subsurface characteristics
beneath individual ARs are also available \citep{Zhao03, Komm08, Jain12, Zhao14, Jain15}.
All these studies converge to a common conclusion that the magnetic regions show
considerably higher subsurface velocities compared to the quiet regions and  the velocities
 increase or decrease with the strength of magnetic elements within the region.
In this study, we find significantly large velocities in ARs as compared to the quiet regions,
however the quantitative measures of these velocity fields depend on various factors 
that are described below. 

In this paper 
we studied two big ARs, NOAA 12192 and 10486,  from two different solar cycles.
There are many similarities as well as differences in the morphology of these regions.
Both were large in size at the time of their appearance on the east limb  and
maintained a complex magnetic configuration, $\beta\gamma\delta$, throughout their
disk passages. Despite being larger in size  and also having
stronger magnetic field than in 10486, the sunspot counts in AR 12192 were 
significantly lower.  In contrast,  AR 10486 had several rotating 
sunspots with high rotational velocity.  These rotating sunspots
are identified by their rotation around the umbral center or other sunspots
within the same AR.  The large  positive-polarity sunspot in this region 
was reported to rotate uniformly at a rate of 2$\degr$.67 hr$^{-1}$ 
for about 46 hours prior to the major flare on 2003 October 28 \citep{Kazachenko10}. 
Studies show that the role of sunspot rotation in flare
energetics is complex and it is often argued that the rotation of the sunspot  
produces energy and magnetic helicity more than the non-rotating case, thus triggering
 large flares. \citet{Kazachenko10} suggested that
the sunspot rotation in AR 10486 contributed significantly to the energy and helicity 
budgets of the whole AR. They further emphasized that the 
shearing motions alone stored sufficient energy and helicity in AR 14086 to account for the 
flare energetics and interplanetary coronal mass ejection helicity content within 
their observational uncertainties.   In our previous study, we investigated
the subsurface flows in AR 11158 that had several rotating sunspots and was the source
region of first X-class flare with a halo CME in cycle 24 \citep{Jain15}. 
The characteristics of subsurface flows in AR 11158 are similar to those in AR 10486 reported in this
paper but 
the amplitude is significantly lower.  A series of CMEs erupted from both
 AR 11158 \citep{Kay17} and AR 10486 \citep{Gopalswamy05}. A closer comparison of flows in AR 10486
(Figure~\ref{flow_2003}) and AR 11158 (Figure 8 of \citet{Jain15}) clearly shows that
the meridional flows are more affected by the CME eruption while rotating sunspots play
a critical role in defining the zonal flows. Note that first CMEs reported from ARs 10486 and 11158
were on 2003 Oct 18 and 2011 Feb 13, respectively. Thus, we obtain similar flow patterns
in these two ARs.  However, the flows in AR 12192 display different 
characteristics, which we interpret in terms of its CME-poor nature. Further,
 by exploiting the high spatial resolution
of HMI, we also studied flows in three sub-regions within AR 11158 and found
that the  leading and trailing polarity regions move faster than the mixed-polarity region.
Since the spatial resolution of GONG Dopplergrams
does not provide sufficient modes near the surface to obtain reliable inferences in individual polarity
regions, our present study is confined to the investigation of subsurface flows in ARs as a whole. 

We show, in Figure~\ref{energy_R2}, the depth variation of flow kinetic energy in
AR 12192 and AR 10486 for 8 consecutive days when several large flares were
triggered. The regions corresponding to major 
flares are also marked in the Figure.   Here, the energy density 
values are calculated using the density profile in the solar interior from model ``S'' 
of \citet{JCD96_modelS}. It can be seen that energy density increases exponentially
with depth, mainly due to the rapidly increasing density. 
Note that these are the resultant energy density values which are elevated
from the quiet region values due to the presence of high magnetic field. 
We notice a significant variation in energy density with time in the case of  
AR 12192 (upper panel) while no significant variation is seen in AR 10486  except on Day 1.
It increases gradually in AR 12192 for the first 5 days although there were two
large flares (M8.7 and X1.6) on Day 4. It appears that the energy released by these flares
 was not sufficient to disrupt the flow which eventually decreased significantly with the
eruption of the X3.1 flare on Day 6. After this event, at a depth of about 13 Mm, almost 50\%
of the energy was dissipated which further decreased from Day 6 to Day 7 and increased from Day 7 to Day 8
with the eruption of several big flares.  
 In all cases, this change in kinetic energy was relatively small in the upper 4 Mm.
However, below this depth, there were significant changes in flow energy - from 400\% around 5.8 Mm to 
650\% around 7.2 Mm, and  this variation gradually decreased in much deeper layers. 
In contrast to these results, the energy variation was minimal 
in AR 10486, which produced several major flares on four consecutive days, including an X1.2 on Day 2,
M6.7 on Day 3, X17.2 on Day 4 and X10.0 on Day 5. We believe that all these flares must have lowered the
kinetic energy, however there were some other processes that rapidly supplied energy to
AR 10486. Thus, the small
variation in flow energy with time  in AR 10486 provided favorable conditions for 
more energetic eruptions, e.g. CMEs in this case. 

 In the next rotation, as shown 
 in Figure~\ref{energy_R3}, the flow energy decreased substantially, almost by a factor
of 4,  in both regions. Further, the flow energy in AR 10508 (10486) is
 significantly higher than in AR 12209 (12192)
in deeper layers while it is comparable in both regions  above 4 Mm.
Our study suggests that additional factors, e.g., the  sunspot rotation combined with the re-organization
of magnetic field in AR 10486, were not sufficient to decrease the flow energy even after several 
large flares and triggered CMEs. Furthermore, in the
absence of sunspot rotation in AR 12192, the re-organization of magnetic field  
 contributed significantly in the substantial release of flow energy  after the  X3.1 flare.

\section{Summary}

AR  12192 is the biggest AR observed in solar cycle 24 
till date. It appeared on the east limb on 2014 October 18 and grew rapidly into
the largest such region since 1990. Composite images of helioseismic farside maps and the direct
frontside observations, in conjunction with STEREO and SDO/AIA EUV observations,
 clearly show that it was a long-lived ARs  that survived  at least 
four CRs and exhibited several unusual phenomena. Over the 
period of four rotations, it had the most
complex magnetic configuration, $\beta\gamma\delta$, during second rotation, i.e in CR 2156
with very strong magnetic field and several large flares without any
associated CMEs.  We measured the
 horizontal subsurface flows in AR 12192 in multiple rotations
 applying the ring-diagram technique to the 
GONG+ Dopplergrams. Our analysis suggests that it had  unusually large horizontal
flow amplitudes before the X3.1 flare on 2014 October 24 which were comparable
to those in AR 10486 during Halloween solar events in 2003.  
 Both regions were located around the same latitude
in southern hemisphere and produced several high M- and X-class flares 
but had different CME productivity. 
 In order to compare the horizontal flows from both active regions, the position of
 ARs in the tracked cubes is crucial. 
 If one region is off centered from the other,
the inferences of mode parameters and the flows are biased by the area
covered. Further, quiet areas within the tile also modifies the results
as inferred quantities are averaged over the entire tile.
To reduce the biasing of these factors, both regions are 
tracked at the Snodgrass rotation rate, which allows us to cover the same area in tracking
ensuring that  the ARs stay at the same location in the tracked cube.
In the case of ARs 12192 and 10486, if one AR is slightly off-centered
from the other AR, both  cover significantly large areas of the tiles and we believe that  
our results are less affected by the positioning of these ARs within the tiles.  

Our analysis further suggests that flow directions in these ARs were
 distinctly different; while meridional flow  in the AR 12192 was poleward  that support
flux transport to the poles, it was equatorward in AR 10486.  
Furthermore, there was a sudden increase  in the magnitude of estimated zonal flow in 
shallow layers in AR 12192 during the X3.1 flare, however, it reversed direction in AR 10486 with 
X17.2 flare. These different flow patterns provided strong twists in 
the horizontal velocity with depth that persisted in AR 10486 throughout the disk passage 
as opposed to AR 12192, which produced a twist only after the eruption of the X3.1 flare
that disappeared soon after. Over the period of eight ring days, we find different flow 
energy patterns in these regions; there was significant variation in flow energy 
with time in the case of  AR 12192  while no significant variation is seen in AR 10486.
It increased gradually in AR 12192 for first four days until the X3.1 was produced and
then declined sharply. We conclude that the sunspot rotation combined with the re-organization
of magnetic field in AR 10486 was not sufficient to decrease  the flow energy even after 
triggering several 
large flares that might have been responsible for CMEs. Furthermore, in the
absence of  rotating sunspots in AR 12192, the re-organization of magnetic field  
contributed significantly to the substantial release of kinetic energy  after the  X3.1 flare.

In future studies, we plan to identify flaring regions with and without
rotating sunspots in both CME-rich and CME-poor categories to obtain a
comprehensive picture linking the role of subsurface flows in the energetic eruptions.
A study based on a large number of dataset may be crucial for providing
subsurface thresholds for the extreme space weather events. It would also be interesting
to investigate the subsurface characteristics of unique regions with CMEs but
 which are not associated with flares.
 
\acknowledgements

We thank the anonymous referee for useful suggestions. This work utilizes GONG data obtained by the NSO Integrated Synoptic Program (NISP), managed by the National Solar Observatory, the Association of Universities for Research in Astronomy (AURA), Inc. under a cooperative agreement with the National Science Foundation. The data were acquired by instruments operated by the Big Bear Solar Observatory, High Altitude Observatory, Learmonth Solar Observatory, Udaipur Solar Observatory, Instituto de Astrof\'{\i}sica de Canarias, and Cerro Tololo Interamerican Observatory. The ring-diagram analysis was carried out using the NSO/GONG ring-diagram pipeline. The STEREO/SECCHI data used here are produced by an international consortium of the Naval Research
Laboratory (USA), Lockheed Martin Solar and Astrophysics Lab (USA), NASA Goddard Space Flight Center (USA) 
Rutherford Appleton Laboratory (UK), University of Birmingham (UK), Max-Planck-Institut fur
Sonnensystemforschung (Germany), Centre Spatiale de Liege (Belgium), Institut d'Optique Theorique et Applique
 (France), Institut d'Astrophysique Spatiale (France). SDO data courtesy of SDO(NASA) and the HMI and AIA
consortium. The farside HMI maps were provided by NASA through the Joint Science Operations Center for the SDO project at Stanford University.

   
\bibliography{Jain_AR12192_10486.bbl}

\begin{thebibliography}{}
\expandafter\ifx\csname natexlab\endcsname\relax\def\natexlab#1{#1}\fi

\bibitem[{{Baldner} \& {Schou}(2012)}]{Baldner12}
{Baldner}, C.~S., \& {Schou}, J. 2012, \apjl, 760, L1

\bibitem[{{Bamba} {et~al.}(2017){Bamba}, {Inoue}, {Kusano}, \&
  {Shiota}}]{Bamba17}
{Bamba}, Y., {Inoue}, S., {Kusano}, K., \& {Shiota}, D. 2017, \apj, 838, 134

\bibitem[{{Basu} {et~al.}(2004){Basu}, {Antia}, \& {Bogart}}]{basu04}
{Basu}, S., {Antia}, H.~M., \& {Bogart}, R.~S. 2004, \apj, 610, 1157

\bibitem[{{Bogart} {et~al.}(2015){Bogart}, {Baldner}, \& {Basu}}]{Bogart15}
{Bogart}, R.~S., {Baldner}, C.~S., \& {Basu}, S. 2015, \apj, 807, 125

\bibitem[{{Chen} {et~al.}(2015){Chen}, {Zhang}, {Ma}, {Yang}, {Li}, {Huang}, \&
  {Xiao}}]{Chen15}
{Chen}, H., {Zhang}, J., {Ma}, S., {et~al.} 2015, \apjl, 808, L24

\bibitem[{{Christensen-Dalsgaard} {et~al.}(1996){Christensen-Dalsgaard},
  {Dappen}, {Ajukov}, {Anderson}, {Antia}, {Basu}, {Baturin}, {Berthomieu},
  {Chaboyer}, {Chitre}, {Cox}, {Demarque}, {Donatowicz}, {Dziembowski},
  {Gabriel}, {Gough}, {Guenther}, {Guzik}, {Harvey}, {Hill}, {Houdek},
  {Iglesias}, {Kosovichev}, {Leibacher}, {Morel}, {Proffitt}, {Provost},
  {Reiter}, {Rhodes}, {Rogers}, {Roxburgh}, {Thompson}, \&
  {Ulrich}}]{JCD96_modelS}
{Christensen-Dalsgaard}, J., {Dappen}, W., {Ajukov}, S.~V., {et~al.} 1996,
  Science, 272, 1286

\bibitem[{{Compagnino} {et~al.}(2017){Compagnino}, {Romano}, \&
  {Zuccarello}}]{Compagnino17}
{Compagnino}, A., {Romano}, P., \& {Zuccarello}, F. 2017, \solphys, 292, 5

\bibitem[{{Gonz{\'a}lez Hern{\'a}ndez} {et~al.}(2007){Gonz{\'a}lez
  Hern{\'a}ndez}, {Hill}, \& {Lindsey}}]{Irene07}
{Gonz{\'a}lez Hern{\'a}ndez}, I., {Hill}, F., \& {Lindsey}, C. 2007, \apj, 669,
  1382

\bibitem[{{Gopalswamy} {et~al.}(2005){Gopalswamy}, {Yashiro}, {Liu},
  {Michalek}, {Vourlidas}, {Kaiser}, \& {Howard}}]{Gopalswamy05}
{Gopalswamy}, N., {Yashiro}, S., {Liu}, Y., {et~al.} 2005, Journal of
  Geophysical Research (Space Physics), 110, A09S15

\bibitem[{{Greer} {et~al.}(2015){Greer}, {Hindman}, {Featherstone}, \&
  {Toomre}}]{Greer15}
{Greer}, B.~J., {Hindman}, B.~W., {Featherstone}, N.~A., \& {Toomre}, J. 2015,
  \apjl, 803, L17

\bibitem[{{Haber} {et~al.}(2002){Haber}, {Hindman}, {Toomre}, {Bogart},
  {Larsen}, \& {Hill}}]{Haber02}
{Haber}, D.~A., {Hindman}, B.~W., {Toomre}, J., {et~al.} 2002, \apj, 570, 855

\bibitem[{{Hill}(1988)}]{Hill88}
{Hill}, F. 1988, \apj, 333, 996

\bibitem[{{Howe} {et~al.}(2015){Howe}, {Komm}, {Baker}, {Harra}, {van
  Driel-Gesztelyi}, \& {Bogart}}]{Howe15}
{Howe}, R., {Komm}, R.~W., {Baker}, D., {et~al.} 2015, \solphys, 290, 3137

\bibitem[{{Inoue} {et~al.}(2016){Inoue}, {Hayashi}, \& {Kusano}}]{Inoue16}
{Inoue}, S., {Hayashi}, K., \& {Kusano}, K. 2016, \apj, 818, 168

\bibitem[{{Jain} {et~al.}(2006){Jain}, {Hill}, {Gonz{\'a}lez Hern{\'a}ndez},
  {Toner}, {Tripathy}, {Armstrong}, \& {Jefferies}}]{Jain06a}
{Jain}, K., {Hill}, F., {Gonz{\'a}lez Hern{\'a}ndez}, I., {et~al.} 2006, in ESA
  Special Publication, Vol. 624, Proceedings of SOHO 18/GONG 2006/HELAS I,
  Beyond the spherical Sun, 127.1

\bibitem[{{Jain} {et~al.}(2008){Jain}, {Hill}, {Tripathy},
  {Gonz{\'a}lez-Hern{\'a}ndez}, {Armstrong}, {Jefferies}, {Rhodes}, \&
  {Rose}}]{Jain08}
{Jain}, K., {Hill}, F., {Tripathy}, S.~C., {et~al.} 2008, in Astronomical
  Society of the Pacific Conference Series, Vol. 383, Subsurface and
  Atmospheric Influences on Solar Activity, ed. {R.~Howe, R.~W.~Komm,
  K.~S.~Balasubramaniam, \& G.~J.~D.~Petrie }, 389

\bibitem[{{Jain} {et~al.}(2012){Jain}, {Komm}, {Gonz{\'a}lez Hern{\'a}ndez},
  {Tripathy}, \& {Hill}}]{Jain12}
{Jain}, K., {Komm}, R.~W., {Gonz{\'a}lez Hern{\'a}ndez}, I., {Tripathy}, S.~C.,
  \& {Hill}, F. 2012, \solphys, 279, 349

\bibitem[{{Jain} {et~al.}(2013{\natexlab{a}}){Jain}, {Tripathy}, {Basu},
  {Bogart}, {Gonz{\'a}lez Hern{\'a}ndez}, {Hill}, \& {Howe}}]{Jain13}
{Jain}, K., {Tripathy}, S., {Basu}, S., {et~al.} 2013{\natexlab{a}}, in Journal
  of Physics Conference Series, Vol. 440, Journal of Physics Conference Series,
  012012

\bibitem[{{Jain} {et~al.}(2013{\natexlab{b}}){Jain}, {Tripathy}, {Basu},
  {Baldner}, {Bogart}, {Hill}, \& {Howe}}]{Jain13a}
{Jain}, K., {Tripathy}, S.~C., {Basu}, S., {et~al.} 2013{\natexlab{b}}, in
  Astronomical Society of the Pacific Conference Series, Vol. 478, Fifty Years
  of Seismology of the Sun and Stars, ed. K.~{Jain}, S.~C. {Tripathy},
  F.~{Hill}, J.~W. {Leibacher}, \& A.~A. {Pevtsov}, 193

\bibitem[{{Jain} {et~al.}(2015){Jain}, {Tripathy}, \& {Hill}}]{Jain15}
{Jain}, K., {Tripathy}, S.~C., \& {Hill}, F. 2015, \apj, 808, 60

\bibitem[{{Jain} {et~al.}(2016){Jain}, {Tripathy}, {Ravindra}, {Komm}, \&
  {Hill}}]{Jain16}
{Jain}, K., {Tripathy}, S.~C., {Ravindra}, B., {Komm}, R., \& {Hill}, F. 2016,
  \apj, 816, 5

\bibitem[{{Jiang} {et~al.}(2016){Jiang}, {Wu}, {Yurchyshyn}, {Wang}, {Feng}, \&
  {Hu}}]{Jiang16}
{Jiang}, C., {Wu}, S.~T., {Yurchyshyn}, V., {et~al.} 2016, \apj, 828, 62

\bibitem[{{Jing} {et~al.}(2015){Jing}, {Xu}, {Lee}, {Nitta}, {Liu}, {Park},
  {Wiegelmann}, \& {Wang}}]{Jing15}
{Jing}, J., {Xu}, Y., {Lee}, J., {et~al.} 2015, Res. Astron. Astrophys., 15,
  1537

\bibitem[{{Kay} {et~al.}(2017){Kay}, {Gopalswamy}, {Xie}, \& {Yashiro}}]{Kay17}
{Kay}, C., {Gopalswamy}, N., {Xie}, H., \& {Yashiro}, S. 2017, \solphys, 292,
  78

\bibitem[{{Kazachenko} {et~al.}(2010){Kazachenko}, {Canfield}, {Longcope}, \&
  {Qiu}}]{Kazachenko10}
{Kazachenko}, M.~D., {Canfield}, R.~C., {Longcope}, D.~W., \& {Qiu}, J. 2010,
  \apj, 722, 1539

\bibitem[{{Komm} {et~al.}(2015){Komm}, {Gonz{\'a}lez Hern{\'a}ndez}, {Howe}, \&
  {Hill}}]{Komm15a}
{Komm}, R., {Gonz{\'a}lez Hern{\'a}ndez}, I., {Howe}, R., \& {Hill}, F. 2015,
  \solphys, 290, 1081

\bibitem[{{Komm} \& {Gosain}(2015)}]{Komm15}
{Komm}, R., \& {Gosain}, S. 2015, \apj, 798, 20

\bibitem[{{Komm} {et~al.}(2012){Komm}, {Howe}, \& {Hill}}]{Komm12}
{Komm}, R., {Howe}, R., \& {Hill}, F. 2012, \solphys, 277, 205

\bibitem[{{Komm} {et~al.}(2008){Komm}, {Morita}, {Howe}, \& {Hill}}]{Komm08}
{Komm}, R., {Morita}, S., {Howe}, R., \& {Hill}, F. 2008, \apj, 672, 1254

\bibitem[{{Liewer} {et~al.}(2014){Liewer}, {Gonz{\'a}lez Hern{\'a}ndez},
  {Hall}, {Lindsey}, \& {Lin}}]{Liewer2014}
{Liewer}, P.~C., {Gonz{\'a}lez Hern{\'a}ndez}, I., {Hall}, J.~R., {Lindsey},
  C., \& {Lin}, X. 2014, \solphys, 289, 3617

\bibitem[{{Liewer} {et~al.}(2012){Liewer}, {Gonz{\'a}lez Hern{\'a}ndez},
  {Hall}, {Thompson}, \& {Misrak}}]{Liewer2012}
{Liewer}, P.~C., {Gonz{\'a}lez Hern{\'a}ndez}, I., {Hall}, J.~R., {Thompson},
  W.~T., \& {Misrak}, A. 2012, \solphys, 281, 3

\bibitem[{{Liu} {et~al.}(2016){Liu}, {Wang}, {Wang}, {Shen}, {Ye}, {Liu},
  {Chen}, {Zhang}, \& {Wang}}]{Liu16}
{Liu}, L., {Wang}, Y., {Wang}, J., {et~al.} 2016, \apj, 826, 119

\bibitem[{{Rabello-Soares} {et~al.}(2016){Rabello-Soares}, {Bogart}, \&
  {Scherrer}}]{Rabello-Soares16}
{Rabello-Soares}, M.~C., {Bogart}, R.~S., \& {Scherrer}, P.~H. 2016, \apj, 827,
  140

\bibitem[{{Rajaguru} {et~al.}(2001){Rajaguru}, {Basu}, \& {Antia}}]{Rajaguru01}
{Rajaguru}, S.~P., {Basu}, S., \& {Antia}, H.~M. 2001, \apj, 563, 410

\bibitem[{{Reinard} {et~al.}(2010){Reinard}, {Henthorn}, {Komm}, \&
  {Hill}}]{Reinard2010}
{Reinard}, A.~A., {Henthorn}, J., {Komm}, R., \& {Hill}, F. 2010, \apjl, 710,
  L121

\bibitem[{{Riley} {et~al.}(2014){Riley}, {Ben-Nun}, {Linker}, {Mikic},
  {Svalgaard}, {Harvey}, {Bertello}, {Hoeksema}, {Liu}, \& {Ulrich}}]{Riley14}
{Riley}, P., {Ben-Nun}, M., {Linker}, J.~A., {et~al.} 2014, \solphys, 289, 769

\bibitem[{{Snodgrass}(1984)}]{snod84}
{Snodgrass}, H.~B. 1984, \solphys, 94, 13

\bibitem[{{Sun} {et~al.}(2015){Sun}, {Bobra}, {Hoeksema}, {Liu}, {Li}, {Shen},
  {Couvidat}, {Norton}, \& {Fisher}}]{Sun15}
{Sun}, X., {Bobra}, M.~G., {Hoeksema}, J.~T., {et~al.} 2015, \apjl, 804, L28

\bibitem[{{Tripathy} {et~al.}(2009){Tripathy}, {Antia}, {Jain}, \&
  {Hill}}]{Tripathy09}
{Tripathy}, S.~C., {Antia}, H.~M., {Jain}, K., \& {Hill}, F. 2009, in
  Astronomical Society of the Pacific Conference Series, Vol. 416,
  Solar-Stellar Dynamos as Revealed by Helio- and Asteroseismology: GONG
  2008/SOHO 21, ed. M.~{Dikpati}, T.~{Arentoft}, I.~{Gonz{\'a}lez
  Hern{\'a}ndez}, C.~{Lindsey}, \& F.~{Hill}, 139

\bibitem[{{Tripathy} {et~al.}(2015){Tripathy}, {Jain}, \& {Hill}}]{Tripathy15}
{Tripathy}, S.~C., {Jain}, K., \& {Hill}, F. 2015, \apj, 812, 20

\bibitem[{{Tripathy} {et~al.}(2012){Tripathy}, {Jain}, {Howe}, {Bogart}, \&
  {Hill}}]{Tripathy12}
{Tripathy}, S.~C., {Jain}, K., {Howe}, R., {Bogart}, R.~S., \& {Hill}, F. 2012,
  Astronomische Nachrichten, 333, 1013

\bibitem[{{Tripathy} {et~al.}(2008){Tripathy}, {Wet}, {Jain}, {Clark}, \&
  {Hill}}]{tripathy2008_CME}
{Tripathy}, S.~C., {Wet}, S., {Jain}, K., {Clark}, R., \& {Hill}, F. 2008, J.
  Astrophys. Astron., 29, 207

\bibitem[{{Yashiro} {et~al.}(2006){Yashiro}, {Akiyama}, {Gopalswamy}, \&
  {Howard}}]{Yashiro06}
{Yashiro}, S., {Akiyama}, S., {Gopalswamy}, N., \& {Howard}, R.~A. 2006, \apjl,
  650, L143

\bibitem[{{Zhao} \& {Kosovichev}(2003)}]{Zhao03}
{Zhao}, J., \& {Kosovichev}, A.~G. 2003, \apj, 591, 446

\bibitem[{{Zhao} {et~al.}(2014){Zhao}, {Kosovichev}, \& {Bogart}}]{Zhao14}
{Zhao}, J., {Kosovichev}, A.~G., \& {Bogart}, R.~S. 2014, \apjl, 789, L7

\bibitem[{{Zhao} {et~al.}(2012){Zhao}, {Nagashima}, {Bogart}, {Kosovichev}, \&
  {Duvall}}]{Zhao12}
{Zhao}, J., {Nagashima}, K., {Bogart}, R.~S., {Kosovichev}, A.~G., \& {Duvall},
  Jr., T.~L. 2012, \apjl, 749, L5

\end{thebibliography}


\clearpage

\clearpage

\clearpage

\clearpage

\clearpage

\clearpage

\clearpage

\clearpage

\clearpage

\clearpage

\clearpage

\clearpage

 \clearpage 

 \clearpage 
 
\setcounter{figure}{13}
\begin{figure}   
   \centerline{
\includegraphics[scale=0.85]{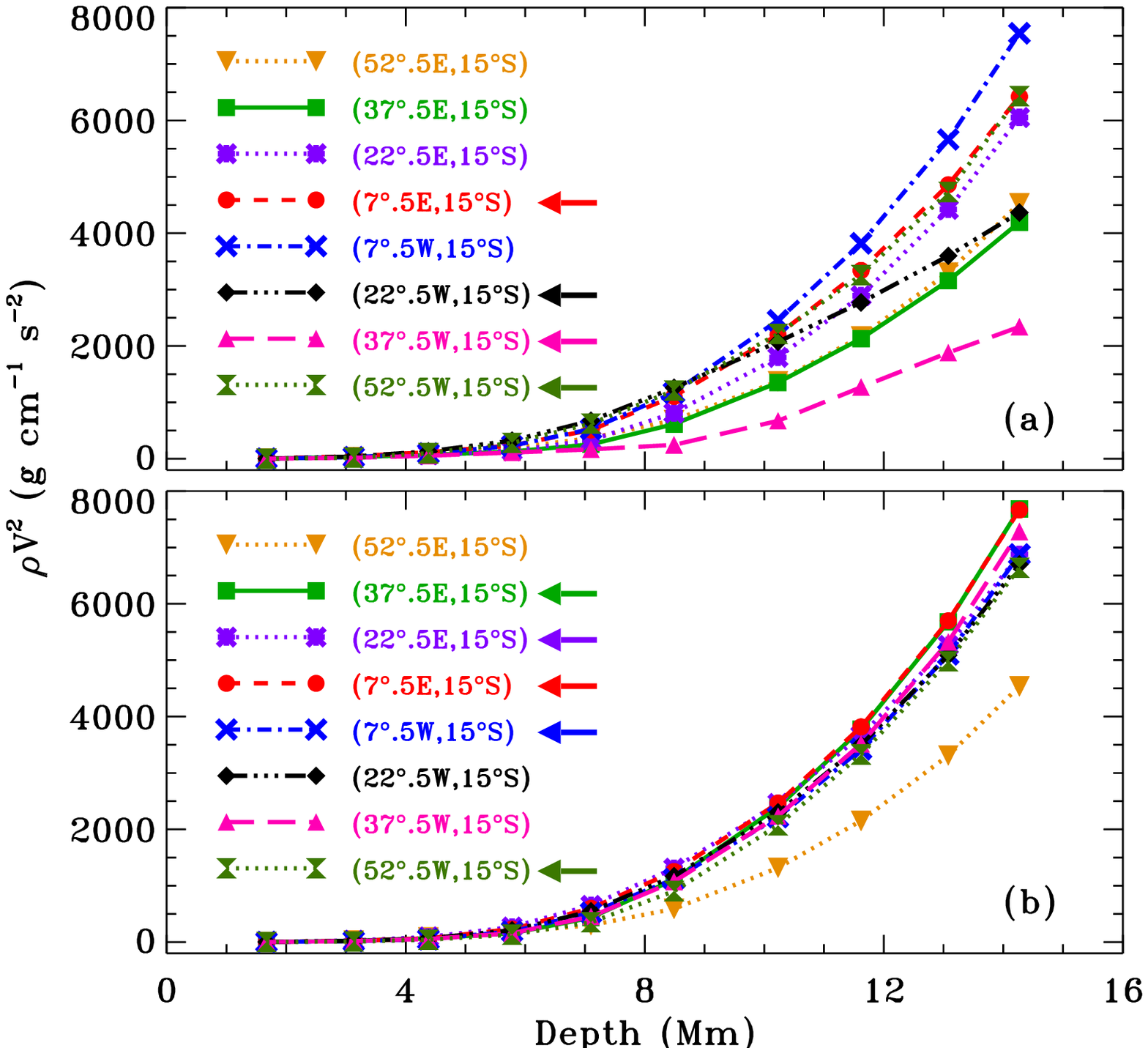}\\
}
            \caption{Depth variation of flow energy density in (a) AR 12192 and
(b) AR 10486 at eight disk locations during as they move across the disk. Both ARs 
produced several X- and high M-class flares (see Tables~\ref{table3} and
\ref{table4}). The regions associated with these flares are marked by arrows.
}
   \label{energy_R2}
   \end{figure}

\begin{figure}   
   \centerline{
\includegraphics[scale=0.85]{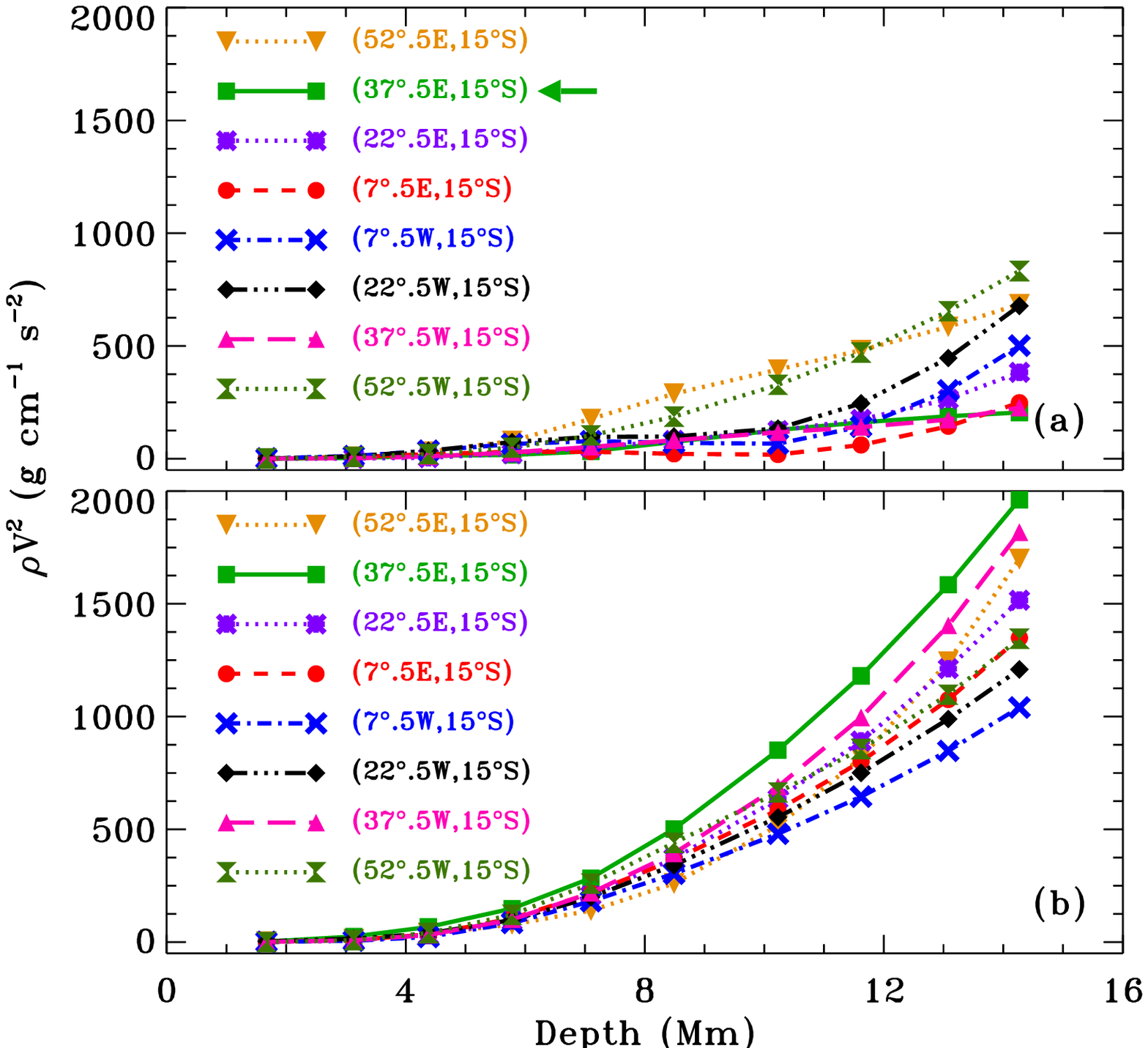}\\
}
            \caption{Depth variation of flow energy density in (a) AR 12209 (12192) and
(b) AR 10508 (10486) eight disk locations during as they move across the disk. These 
ARs reappeared on the Earth-side limb after completing
the Sun's rotation and their NOAA
numbers in the previous rotation are given in parenthesis. The AR 12209
produced a high M-class flare (see Table~\ref{table3}) 
and the region associated with this flare is marked by the arrow.
}
   \label{energy_R3}
   \end{figure}

\clearpage   
\begin{deluxetable}{lcccc}
\tabletypesize{\scriptsize} 
\tablecaption{Evolution of AR 12192 in multiple CRs. Note that different NOAA numbers were assigned to
this region  in each CR.\tablenotemark{a}   \label{table1} }
\tablewidth{0pt}
\tablehead{
\colhead{Carrington}  &\colhead{Assigned}  & \multicolumn{3}{c}{Number of Flares}\\
\cline{3-5}
\colhead{Rotation}   & \colhead{AR Number} &          \colhead{C}  & \colhead{M}  & \colhead{X}
}
\startdata
2155  & 12173 & 19 & 2 & 0\\
{\bf 2156}  & {\bf 12192} & {\bf 73} & {\bf 35} &{\bf  6} \\
2157 & 12209  & 33 & 3 & 0\\
2158 & 12237 & 3 & 0 & 0\\
\enddata
\tablecomments{}
\tablenotetext{a}{Source: NOAA/SWPC}
\end{deluxetable}

\clearpage  
\begin{deluxetable}{rcccccccccccc}
\tabletypesize{\scriptsize} 
\tablecaption{Mean Magnetic Activity Index (MAI) and Standard Deviation (STDDEV) for Quiet Regions.\label{table2}  }
\tablewidth{0pt}
\tablehead{
\colhead{Reference Longitude}&                 
\colhead{}          & \multicolumn{2}{c}{R1}& 
\colhead{}          & \multicolumn{2}{c}{R2}&    
\colhead{}          & \multicolumn{2}{c}{R3}& 
\colhead{}          & \multicolumn{2}{c}{R4}\\
\cline{3-4} \cline{6-7} \cline{9-10} \cline{12-13}
 \colhead{ }& \colhead{}& \colhead{MAI}& \colhead{STDDEV}& 
                      \colhead{}& \colhead{MAI}& \colhead{STDDEV}&
                      \colhead{}& \colhead{MAI}& \colhead{STDDEV}& 
                      \colhead{}& \colhead{MAI}& \colhead{STDDEV}\\
 \colhead{(Degree)}&  \colhead{}& \colhead{(G)}& \colhead{(\%)}& 
                      \colhead{}& \colhead{(G)}& \colhead{(\%)}& 
                      \colhead{}& \colhead{(G)}& \colhead{(\%)}& 
                      \colhead{}& \colhead{(G)}& \colhead{(\%)}
}
\startdata
$-$52.5 &  &   0.32 &   15.6&&	     0.35  &  17.1&&	     0.29  &  17.2	& &    0.35  &  17.1\\
 $-$45.0& &    0.38 &   13.1&&	     0.42  &  14.3&&	     0.39   & 12.8	& &    0.39  &  12.8\\
$-$37.5 &&     0.43 &   11.6&&	     0.47  &  10.6&&	     0.45  &  11.1	& &    0.44  &  13.6\\
$-$30.0 &&     0.48 &   10.4&&	     0.50  &  10.0&&	     0.50 &   10.0	& &    0.43  &  11.6\\
$-$22.5 &&     0.50 &   10.0&&	     0.54  &  9.2&&	     0.52 &   9.6	& &    0.44  &  11.3\\
 $-$15.0&&     0.49 &   10.2&&	     0.55   & 9.0&&	     0.49 &   10.2	& &    0.48  &  10.4\\
 $-$7.5& &    0.53 &   9.4&&         0.60  & 8.3 &&	     0.54 &   9.2	&&     0.53  &  9.4\\
   0& &    0.53 &   9.4&&           0.58  &  8.6&&	     0.55 &   9.0	& &    0.55  &  9.1\\
  7.5&&    0.53 &   9.4&&            0.58  &  8.6&&	     0.54&    9.2	& &    0.52  &  9.6\\
 15.0&&     0.51 &   9.8&&	     0.57  &  8.8&&	     0.56 &   8.9	&&     0.53  &  11.3\\
 22.5&&     0.50 &   10.0&&	     0.54  &  9.2&&	     0.53 &   9.4	& &    0.51  &  11.7\\
 30.0& &    0.49 &   12.2&&	     0.50  &  9.0&&	     0.51 &   9.8	& &    0.50  &  12.0\\
 37.5& &    0.42 &   14.3&&	     0.44  &  13.6&&	     0.48 &   10.4	 & &   0.49  &  12.2\\
  45.0& &   0.36 &   13.9&&	     0.40  &  15.0&&	     0.42 &   11.9	 & &   0.47   & 12.7\\
 52.5&&     0.31&    16.1& &	     0.32  &  18.7&&	     0.33 &   18.1	 & &   0.38  &  15.8\\
\enddata
\tablecomments {Values are given at 15$^{\degr}$S and 15 different
locations across the disk for four  ranges of  $B_0$; R1: 7$^{\degr}$.05 -- 6$^{\degr}$.51,
 R2: 5$^{\degr}$.31 -- 3$^{\degr}$.99, 
R3: 2$^{\degr}$.58 -- 0$^{\degr}$.91, and R4: $-$0$^{\degr}$.69 -- $-$2$^{\degr}$.39. 
These ranges of  $B_0$ correspond to the data obtained from 2008 September to 2008 December.}
\end{deluxetable}

\clearpage

\begin{deluxetable}{lrccrrrrrcc}
\tabletypesize{\scriptsize} 
\tablecaption{Various regions representing AR 12192 in four rotations.   \label{table3}  }
\tablewidth{0pt}
\tablehead{
 \colhead{CR}              & \colhead{AR}                & \colhead{QR} & 
 \colhead{Day}             & \colhead{}                    & \multicolumn{2}{c} {Timeseries} & 
 \colhead{}                & \multicolumn{2}{c} {Location} & \colhead{DC}\\
\cline{6-7} \cline{9-10}
 \colhead{}                & \colhead{}                    & \colhead{}    &
 \colhead { \#}             & \colhead{}                    & \colhead{Start} & \colhead {End}             & 
                              \colhead{}                    & \colhead{Long} & \colhead{Lat}              & 
\colhead{ (\%)}
}
\startdata
 2155& 12172& R1&&  & & &&& &  \\
&&&  1&& 2014 Sep 21 16:28UT& 2014 Sep 22 20:11UT& &  $-$52$^{\mathrm{o}}$.5 &   $-$15$^{\mathrm{o}}$&  86\\
&&&  2&& 2014 Sep 22 19:44UT& 2014 Sep 23 23:27UT&&  $ -$37$^{\mathrm{o}}$.5 &   $-$15$^{\mathrm{o}}$&  92\\
&&& 3&& 2014 Sep 23 23:01UT& 2014 Sep 25 02:44UT&&  $ -$22$^{\mathrm{o}}$.5 & $-$15$^{\mathrm{o}}$& 98 \\
&&&  4&& 2014 Sep 2502:17UT& 2014 Sep 26 06:00UT& &   $ -$7$^{\mathrm{o}}$.5  & $-$15$^{\mathrm{o}}$& 89 \\
&&&  5&& 2014 Sep 26 05:34UT& 2014 Sep 27 09:17UT&&    7$^{\mathrm{o}}$.5 & $-$15$^{\mathrm{o}}$& 84 \\
&&& 6&& 2014 Sep 27 08:51UT& 2014 Sep 28 12:34UT& &   22$^{\mathrm{o}}$.5  & $-$15$^{\mathrm{o}}$& 94 \\
&&& 7&& 2014 Sep 28 12:08UT  & 2014 Sep 29 15:51UT & &  37$^{\mathrm{o}}$.5 & $-$15$^{\mathrm{o}}$& 77 \\
&&& 8&& 2014 Sep 29 15:24UT  & 2014 Sep 30 19:07UT & &  52$^{\mathrm{o}}$.5 & $-$15$^{\mathrm{o}}$& 81 \\
\hline
 {\bf2156}&{\bf12192}&R2& &&  & & & &&    \\
&&&  1& &2014 Oct 18 23:18UT & 2014 Oct 20 03:01UT& &  $-$52$^{\mathrm{o}}$.5 &   $-$15$^{\mathrm{o}}$& 79 \\
&&&  2& &2014 Oct 20 02:36UT & 2014 Oct 21 06:19UT&&  $ -$37$^{\mathrm{o}}$.5 &   $-$15$^{\mathrm{o}}$& 97 \\
&&& 3&& 2014 Oct 21 05:53UT& 2014 Oct 22 09:36UT&&   $-$22$^{\mathrm{o}}$.5 & $-$15$^{\mathrm{o}}$& 96 \\
&&& 4& &\tablenotemark{a}{\bf2014 Oct 22 09:11UT}&  {\bf2014 Oct 23 12:54UT}& &  {\bf$-7$$^{\mathrm{o}}$.5}  & {\bf$-$15$^{\mathrm{o}}$}&  {\bf 70} \\
&&& 5& & 2014 Oct 23 12:29UT& 2014 Oct 24 16:12UT&&    7$^{\mathrm{o}}$.5 & $-$15$^{\mathrm{o}}$&  89\\
&&& 6& & \tablenotemark{b}{\bf2014 Oct 24 15:47UT}&  {\bf2014 Oct 25 19:30UT}& & {\bf22$^{\mathrm{o}}$.5}  & {\bf$-$15$^{\mathrm{o}}$}&   {\bf 88}\\
&&& 7& & \tablenotemark{c}{\bf2014 Oct 25 19:05UT}&   {\bf2014 Oct 26 22:48UT}&&  {\bf37$^{\mathrm{o}}$.5} & {\bf$-$15$^{\mathrm{o}}$} &  {\bf 78}\\
&&&  8& &\tablenotemark{d}{\bf2014 Oct 26 22:22UT} &  {\bf2014 Oct 28 02:05UT}&&  {\bf52$^{\mathrm{o}}$.5} & {\bf$-$15$^{\mathrm{o}}$}& {\bf86} \\
\hline
 2157& 12209& R3& & & & &&&&    \\
&&& 1& &2014 Nov 15 06:32UT& 2014 Nov 16 10:15UT&&   $-$52$^{\mathrm{o}}$.5 & 15$^{\mathrm{o}}$&  77\\
& && 2& &\tablenotemark{e}{\bf2014 Nov 16 09:51UT}&  {\bf2014 Nov  17 13:34UT}& & {\bf$-$37$^{\mathrm{o}}$.5} &   {\bf$-$15$^{\mathrm{o}}$}&  {\bf 93} \\
&&& 3& &2014 Nov 17 13:09UT& 2014 Nov 18 16:52UT&&   $-$22$^{\mathrm{o}}$.5 & $-$15$^{\mathrm{o}}$&  97\\
&&& 4& &2014 Nov 18 16:28UT& 2014 Nov 19  20:11UT&&    $ -$7$^{\mathrm{o}}$.5  & $-$15$^{\mathrm{o}}$& 94\\
&&& 5& &2014 Nov 19 19:46UT& 2014 Nov 20 23:29UT&&   7$^{\mathrm{o}}$.5 & $-$15$^{\mathrm{o}}$& 91 \\
&&&6& &2014 Nov 20  23:05UT& 2014 Nov 22 02:48UT&&    22$^{\mathrm{o}}$.5  & $-$15$^{\mathrm{o}}$& 92 \\
&&&7& &2014 Nov 22 02:24UT& 2014 Nov 23 06:07UT& &  37$^{\mathrm{o}}$.5 & $-$15$^{\mathrm{o}}$& 94 \\
&&&8& &2014 Nov 23 05:43UT& 2014 Nov 24 09:26UT& &  52$^{\mathrm{o}}$.5 & $-$15$^{\mathrm{o}}$& 96 \\
\hline
 2158& 12237& R4& & & & &&& &   \\
& &&1& &2014 Dec 12 14:07UT& 2014 Dec 13 17:50UT&&   $-$52$^{\mathrm{o}}$.5 &   $-$15$^{\mathrm{o}}$& 79 \\
& &&2& &2014 Dec 13 17:26UT& 2014 Dec 14 21:09UT&&   $-$37$^{\mathrm{o}}$.5 &   $-$15$^{\mathrm{o}}$& 83 \\
&&&3&& 2014 Dec 14 20:46UT& 2014 Dec 16 00:29UT& &  $-$22$^{\mathrm{o}}$.5 & $-$15$^{\mathrm{o}}$& 88 \\
&&& 4&& 2014 Dec 16 00:05UT& 2014 Dec 17 03:48UT& &    $-$7$^{\mathrm{o}}$.5  & $-$15$^{\mathrm{o}}$& 96 \\
&&& 5&& 2014 Dec 17 03:25UT& 2014 Dec 18 07:08UT&&   7$^{\mathrm{o}}$.5 & $-$15$^{\mathrm{o}}$& 97 \\
&&&6&& 2014 Dec 18 06:44UT& 2014 Dec 19 10:27UT&&    22$^{\mathrm{o}}$.5  & $-$15$^{\mathrm{o}}$& 94 \\
&&&7&& 2014 Dec 19 10:04UT& 2014 Dec 20 13:47UT&&    37$^{\mathrm{o}}$.5 & $-$15$^{\mathrm{o}}$& 93 \\
&&&8&& 2014 Dec 20 13:23UT& 2014 Dec 21 17:06UT& &   52$^{\mathrm{o}}$.5 & $-$15$^{\mathrm{o}}$& 75 \\
\enddata
\\
\tablecomments {The analysis is carried out 
using 1664 60-s Dopplergrams at each disk location. The locations are listed for the reference
image which is the center image  in each timeseries. Corresponding quiet region (QR) sets used in individual 
CRs  and the duty cycle (DC) for each timeseries are also given.   Regions associated with 
major M- ($\ge$ 5.0) and X- class flares are listed in bold.\\ 
$^{a}$ M8.7 (Oct 22: 01:16UT -- 02:28UT, X1.6 (Oct 22: 14:02UT -- 22:30UT)\\
 $^{b}$ X3.1 (Oct 24: 20:50UT -- 00:14UT), X1.0 (Oct 25: 16:55UT -- 18:11UT)\\ 
$^{c}$ X2.0 (Oct 26: 10:04UT -- 11:18UT)\\ 
$^{d}$ M7.1 (Oct 27: 00:01UT -- 10:22UT), M6.7 (Oct 27: 09:59UT -- 10:26UT), X2.0 (Oct 27: 14:04UT -- 15:31UT)\\
$^{e}$ M5.7 (Nov 16: 17:35UT -- 17:57UT).}
\end{deluxetable}

\clearpage

\begin{deluxetable}{lrccrrrrrcc}
\tabletypesize{\scriptsize} 
\tablecaption{Various regions representing AR 10486 in two rotations.  \label{table4}  }
\tablewidth{0pt}
\tablehead{
 \colhead{CR}              & \colhead{AR}                & \colhead{QR} & 
 \colhead{Day}             & \colhead{}                    & \multicolumn{2}{c} {Timeseries} & 
 \colhead{}                & \multicolumn{2}{c} {Location} & \colhead{DC}\\
\cline{6-7} \cline{9-10}
 \colhead{}                & \colhead{}                    & \colhead{}    &
 \colhead { \#}             & \colhead{}                    & \colhead{Start} & \colhead {End}         & 
                           \colhead{}                    & \colhead{Long} & \colhead{Lat}           & 
\colhead{ (\%)}
}
\startdata
{\bf2009}&{\bf10486}&R2& & & & &&&&\\
&&&1&&2003 Oct 24 15:58UT& 2003 Oct 25 19:41UT& &$ $-52$^{\mathrm{o}}$.5 & $-$15$^{\mathrm{o}}$&96\\
&&& 2&&\tablenotemark{a}{\bf2003 Oct 25 19:16UT}&  {\bf2003 Oct 26 22:59UT}&&  {\bf$-$37$^{\mathrm{o}}$.5} &   {\bf$-$15$^{\mathrm{o}}$}&{\bf73}\\
&&&3&&\tablenotemark{b}{\bf2003 Oct 26 22:34UT}& {\bf2003 Oct 28 02:17UT}&&  {\bf  $-$22$^{\mathrm{o}}$.5 }&  {\bf  $-$15$^{\mathrm{o}}$}&{\bf88}\\
&&&4&&  \tablenotemark{c}{\bf2003 Oct 28 01:50UT}&{\bf2003 Oct  29 05:33UT}&& {\bf  $-$7$^{\mathrm{o}}$.5  }& {\bf$-$15$^{\mathrm{o}}$}&{\bf74}\\
&&& 5&& \tablenotemark{d}{\bf2003 Oct 29 05:08UT}& {\bf2003 Oct 30 08:51UT}&&  {\bf   7$^{\mathrm{o}}$.5} & {\bf  $-$15$^{\mathrm{o}}$}&{\bf91}\\
&&&6&&2003 Oct  30 08:26UT& 2003 Oct  31 12:09UT& & 22$^{\mathrm{o}}$.5  & $-$15$^{\mathrm{o}}$&100\\
&&&7&&2003 Oct 31 11:44UT& 2003 Nov 01 15:27UT& & 37$^{\mathrm{o}}$.5 & $-$15$^{\mathrm{o}}$&93\\
&&&8&&\tablenotemark{e}{\bf2003 Nov 01 15:02UT}& {\bf2003 Nov 02 18:45UT}& & {\bf52$^{\mathrm{o}}$.5 }& {\bf$-$15$^{\mathrm{o}}$}&{\bf95}\\

\hline
2010&10508&R3& & & & &&&&\\
&&&1&&2003 Nov 20 23:22UT& 2003 Nov 21 03:05UT& & $-$52$^{\mathrm{o}}$.5 & $-$15$^{\mathrm{o}}$&96\\
& && 2&&2003 Nov 22 02:40UT& 2003 Nov 23 06:23UT&&  $-$37$^{\mathrm{o}}$.5 &  $-$ 15$^{\mathrm{o}}$&91\\
&&& 3&&2003 Nov 23 05:58UT&2003 Nov 24 09:41UT& & $-$22$^{\mathrm{o}}$.5 & $-$15$^{\mathrm{o}}$&91\\
&&& 4&&2003 Nov 24 09:16UT& 2003 Nov  25 12:59UT&&    $-$7$^{\mathrm{o}}$.5  & $-$15$^{\mathrm{o}}$&84\\
&&& 5&&2003 Nov 25 12:34UT& 2003 Nov 26 16:17UT& & 7$^{\mathrm{o}}$.5 & $-$15$^{\mathrm{o}}$&93\\
&&&6&&2003 Nov  26 15:52UT& 2003 Nov  27 19:35UT& &  22$^{\mathrm{o}}$.5  &$-$ 15$^{\mathrm{o}}$&100\\
&&&7&&2003 Nov 27 19:10UT& 2003 Nov 28 22:53UT&&  37$^{\mathrm{o}}$.5 & $-$15$^{\mathrm{o}}$&96\\
&&&8&&2003 Nov 28 22:28UT& 2003 Nov 30 02:11UT&&  52$^{\mathrm{o}}$.5 &$-$ 15$^{\mathrm{o}}$&95\\
\enddata
\\
\tablecomments {The analysis is carried out 
using 1664 60-s Dopplergrams at each disk location. The locations are listed for the reference
image which is the center image  in each timeseries. Corresponding quiet region (QR) sets used in individual 
CRs  and the duty cycle (DC) for each timeseries are also given.   Regions associated with 
major M- ($\ge$ 5.0) and X- class flares are listed in bold.\\ 
$^{a}$ X1.2(Oct 26: 06:17UT -- )\\
$^{b}$ M6.7 (Oct 27: 12:51UT  -13:04UT)\\
$^{c}$ X17.2 (Oct 28: 10:01UT -- 14:20UT)\\
$^{d}$ X10.0 (Oct 29: 20:37UT -- 22:53UT)\\
$^{e}$ X8.3 (Nov 02: 17:04UT -- 19:54UT).}
\end{deluxetable}

\end{document}